\documentclass[11pt]{article}

\usepackage{fullpage}

\usepackage{latexsym}
\usepackage{amsmath}
\usepackage{amssymb}
\usepackage{mathtools}
\usepackage{url}
\usepackage{xspace}
\usepackage{color}
\usepackage[usenames,dvipsnames,table]{xcolor}
\usepackage{tikz}
\usepackage{dirtytalk}
\usepackage{placeins}
\usepackage{xcolor}

\def\CC{{C\nolinebreak[4]\hspace{-.05em}\raisebox{.4ex}{\tiny\textbf{++}}}}

\usetikzlibrary{positioning,decorations,decorations.pathmorphing,arrows,arrows.meta,backgrounds,fit,patterns}

\newcommand{\eg}{e.g.\@\xspace}
\newcommand{\ie}{i.e.\@\xspace}

\newcommand{\dist}{\mathrm{dist}}

\newcommand{\DIJK}{\mathrm{DIJK}}
\newcommand{\IN}{\mathrm{IN}}
\newcommand{\INSIMPLE}{\mathrm{IN\-SIMP\-LE}}
\newcommand{\INSTATIC}{\mathrm{IN\-STA\-TIC}}
\newcommand{\OUT}{\mathrm{OUT}}
\newcommand{\OUTWEAK}{\mathrm{OUT\-WEAK}}
\newcommand{\OUTSIMPLE}{\mathrm{OUT\-SIMP\-LE}}
\newcommand{\OUTSTATIC}{\mathrm{OUT\-STA\-TIC}}
\newcommand{\ORACLE}{\mathrm{ORACLE}}

\newcommand{\Qdelmin}{\textsf{delete-min}\xspace}
\newcommand{\Qfindmin}{\textsf{find-min}\xspace}
\newcommand{\Qdecrease}{\textsf{decrease-key}\xspace}
\newcommand{\Qincrease}{\textsf{increase-key}\xspace}
\newcommand{\Qdelete}{\textsf{delete}\xspace}
\newcommand{\Qinsert}{\textsf{insert}\xspace}
\newcommand{\Qbuild}{\textsf{build}\xspace}

\newcommand{\InF}{\mathrm{InF}}
\newcommand{\InU}{\mathrm{InU}}
\newcommand{\OutF}{\mathrm{OutF}}
\newcommand{\OutU}{\mathrm{OutU}}
\newcommand{\InCrit}{\mathrm{InCrit}}
\newcommand{\OutCrit}{\mathrm{OutCrit}}

\newtheorem{lemma}{Lemma}
\newtheorem{definition}{Definition}
\newtheorem{proposition}{Proposition}
\newenvironment{proof}{\emph{Proof:}}{$\Box$\newline}

\title{More Parallelism in Dijkstra's Single-Source Shortest Path Algorithm}
\author{Michael Kainer, Jesper Larsson Tr\"aff\\
TU Wien (Vienna University of Technology), Faculty of Informatics \\
Vienna, Austria\\
email: \url{traff@par.tuwien.ac.at}
}

\begin{document}
\maketitle

\begin{abstract}
  Dijkstra's algorithm for the Single-Source Shortest Path (SSSP)
  problem is notoriously hard to parallelize in $o(n)$ depth, $n$
  being the number of vertices in the input graph, without increasing
  the required parallel work unreasonably. Crauser et al.\ (1998)
  presented observations that allow to identify more than a single
  vertex at a time as correct and correspondingly more edges to be
  relaxed simultaneously. Their algorithm runs in parallel phases, and
  for certain random graphs they showed that the number of phases is
  $O(n^{1/3})$ with high probability. A work-efficient CRCW PRAM with
  this depth was given, but no implementation on a real, parallel
  system.

  In this paper we strengthen the criteria of Crauser et al., and
  discuss tradeoffs between work and number of phases in their
  implementation. We present simulation results with a range of common
  input graphs for the depth that an ideal parallel algorithm that can
  apply the criteria at no cost and parallelize relaxations without
  conflicts can achieve. These results show that the number of phases
  is indeed a small root of $n$, but still off from the shortest path
  length lower bound that can also be computed.

  We give a shared-memory parallel implementation of the most
  work-efficient version of a Dijkstra's algorithm running in parallel
  phases, which we compare to an own implementation of the well-known
  $\Delta$-stepping algorithm. We can show that the work-efficient
  SSSP algorithm applying the criteria of Crauser et al. is
  competitive to and often better than $\Delta$-stepping on our chosen
  input graphs. Despite not providing an $o(n)$ guarantee on the
  number of required phases, criteria allowing concurrent relaxation
  of many correct vertices may be a viable approach to practically
  fast, parallel SSSP implementations.
\end{abstract}

\section{Introduction}

The single-source shortest path (SSSP) problem is one of the most
productive problems in computer science. The SSSP problem has so far proven
hard to parallelize, and no algorithms with linear, parallel speedup
for the general case are so far known.  For graphs with non-negative
edge costs, Dijkstra's algorithm~\cite{Dijkstra59} (or variations
thereof) is practically and theoretically attractive, but
unfortunately also strictly sequential: The vertices of the graph are
processed (identified as correct) one after the other, with possible
parallelism only in the relaxation step. Better sequential bounds than
possible with Dijkstra's algorithm are known for the RAM
model~\cite{Hagerup00:sssp}, especially for undirected
graphs~\cite{Thorup99}, and imply that vertices are processed in a
different order than that implied by Dijkstra's algorithm, but such
algorithms are hardly practical. Parallelizations of these algorithms
are also not known. Zwick~\cite{Zwick01} gives an excellent overview of
approaches to the SSSP and related problems.

In this paper~\cite{Kainer18}\footnote{The paper is based on the
  Master's thesis of Michael Kainer, commenced early 2018 and completed
  November 2018.}, we explore criteria that allow Dijkstra's algorithm
to settle more than a single vertex at a time, thus providing
potentially more parallelism by allowing relaxation of the edges of
several vertices in the same parallel phase. Such criteria for
identifying whether a vertex is already correct (a shortest path
found) were presented explicitly by Crauser et
al.~\cite{CrauserMehlhornMeyerSanders98}. Concretely, they proposed
(and combined) two such criteria leading to a version of Dijkstra's
algorithm that runs in parallel phases.  In each phase, at least one,
but possibly much more vertices are identified as already correct
based on the tentative distances computed so far. In a phase, all
correct vertices are settled, and all edge relaxations potentially
done in parallel. Crauser et al.\ analyzed the expected number of
phases for certain random graphs.  For random graphs with $n$
vertices, their combined criteria reduce the number of phases to
$O(n^{1/3})$ with high probability. Crauser et al.\ gave a CRCW PRAM
implementation of their algorithm, and discussed simulation results
confirming the analytical results. However, they did not give any real
implementations.

In this paper, we take up on the work of Crauser et al. We first give
much stronger criteria exploiting in each phase more information
available in the explored part of the graph as well as in the
unexplored part, and discuss the work required in order to decide
efficiently whether a vertex is correct. We present a more extensive
simulation study with different types of graphs, and compare the
number of phases to the lower bound on the number of phases that can
also be computed. This study shows that the strengthened criteria
improve over Crauser et al.'s criteria by reducing the number of
phases to a smaller root of the number of vertices, unfortunately at
the drawback of being more expensive to evaluate. We give an
implementation for shared-memory multi-core processors of the Crauser
et al.\ criteria, and compare running time and scalability (speed-up)
to what can be achieved with $\Delta$-stepping~\cite{MeyerSanders03}.
Our benchmarks show that the SSSP algorithm running in phases and
using only Crauser's et al.'s criteria is an attractive alternative to
$\Delta$-stepping, in many cases providing significantly higher
speed-up, and never performing worse. This approach to parallelizing
Dijkstra's SSSP algorithm clearly merits further attention. It is
important to point out, though, that the criteria for identifying
correct vertices described here (and by Crauser et al.) do not lead to
any worst-case guarantees on the number of phases being strictly
smaller than the number of vertices or any other theoretical
improvements of Dijkstra's SSSP algorithm, but they do provide much
room for engineering the implementations, and as our experiments show
for many types of graphs do lead to very significant reductions in the
number of phases.

Criteria to settle vertices early have been applied in several
algorithms that achieve average-case, (almost) linear, sequential
running time for the SSSP problem, see,
\eg,~\cite{Goldberg08,Hagerup06,Meyer03}. Interestingly, these papers
exploit only variants of the $\IN$-criteria discussed in
Section~\ref{sec:criteria}, under different names like ``Caliber
Lemma'' by Goldberg~\cite{Goldberg08}.  Recently, Garg~\cite{Garg18}
also took up on the idea of Crauser et al., and presented improvements
and implementation ideas, in particular in order to reduce the number
of expensive priority queue operations which can benefit a sequential
implementation. Garg's first two criteria are weaker than the
strongest of the criteria we discuss here in the sense of potentially
identifying fewer vertices as correct, but may lead to a cheaper
implementation. Garg also explores an orthogonal idea with his third
criteria of maintaining lower bounds on the vertex distances.
Like~\cite{CrauserMehlhornMeyerSanders98}, Garg's paper contains no
real implementation, and also no simulation results investigating the
strength of the various criteria he discusses.  It was an eerie
coincidence that our work was done completely independently of
Garg~\cite{Garg18} and vice versa.  The $\Delta$-stepping idea is
extended by Blelloch et al.~\cite{BlellochGuSunTangwongsan16} with a
careful work-depth tradeoff analysis. The paper gives a simulation
based experimental analysis of the bounds, but presents no real
implementation of the ideas.

\section{Preliminaries}

Let $G=(V,E)$ be a \emph{directed graph} with vertices $V$ and edges
$E$, with a \emph{cost function} $c:E\rightarrow\mathbb{R}_{\geq 0}$
assigning a non-negative, real-valued \emph{cost} to each edge of
$G$. The \emph{cost of a path} is the sum of the costs of the edges along the
path, that is $c(P) = \sum_{e\in P}c(e)$ for path
$P=[(u_0,u_1),(u_1,u_2),\ldots (u_{p-1},u_p)]$ consisting of edges
$e_i=(u_i,u_{i+1}),0\leq i<p$. The \emph{length of a path} is the number
of edges along the path, that is $p$ for the path $P$.  Since edge
costs are non-negative, for each vertex $u$ and each vertex $v$
reachable from $u$ in $G$ there is a path having the smallest cost
over all possible paths from $u$ to $v$.  Such a path is called a
\emph{shortest path} from $u$ to $v$, and $\dist(u,v)$ is defined to
be the cost of a shortest path from $u$ to $v$, with
$\dist(u,u)=0$ (empty path; for convenience we also define
$\dist(u,v)=\infty$ if $v$ is not reachable from $u$). Let $s\in V$ be
a given \emph{source} vertex. The \emph{Single-Source Shortest Path
  problem} (SSSP) is to compute $\dist(s,u)$ for each $u\in V$.

Dijkstra's algorithm (implicitly) maintains a partition of the
vertices of $G$ into \emph{settled} vertices $S$, \emph{fringe}
vertices $F$, and \emph{unexplored} vertices $U$, and works as
follows. A \emph{tentative distance} $d[u]$ is associated with each
vertex $u\in V$. The settled vertices $u\in S$ have the property that
$d[u]=\dist(s,u)$. For vertices $u\in F$, $d[u]\geq\dist(s,u)$, and
additionally $d[u]$ is the cost of a shortest path from $s$ to $u$
passing only through vertices in $S$. For vertices $u\in U$, no such
path through only vertices in $S$ exists. To establish the invariants,
Dijkstra's algorithm initially sets $S=\emptyset, F=\{s\}$ with
$d[s]=0$, and $U=V\setminus F$. The crucial observation is that for a
vertex $u\in F$ with $d[u] = \min_{v\in F}d[v]$, it holds that
$d[u]=\dist(s,u)$, such that $u$ can be \emph{settled}: Vertex $u$ is
moved to the set $S$. To reestablish the invariants, all outgoing
edges incident to $u$ are \emph{relaxed}: For each edge $(u,v)$, if
$v\in F$ and $d[u]+c(u,v)<d[v]$, $d[v]$ is updated to the shorter path
cost $d[u]+c(u,v)$ obtained by passing through $u$, and if $v\in U$,
$d[v]$ is set to $d[u]+c(u,v)$ and $v$ is moved to $F$ from $U$ (if
$v\in S$, there is nothing to be done, since $d[u]+c(u,v)\geq d[v]$).

In the following, we say that a vertex $u\in F$ is \emph{correct} if
$d[u]=\dist(s,u)$. Dijkstra's algorithm identifies and settles one
correct vertex per iteration.

Now let $n=|V|$ be the number of vertices and $m=|E|$ the number of
edges of $G$. It is well-known, but non-trivial that with the right
graph representation and data structures, Dijkstra's algorithm can be
implemented to run in $O(n\log n+m)$ operations~\cite{FredmanTarjan87}.
An overview of priority queues as needed for Dijkstra's algorithm can
be found in~\cite{Brodal13}.

The algorithm is strictly sequential, since correct vertices are
identified one after the other, and only possibly the edge relaxation
can be done in parallel. Such a straightforward parallelization of
Dijkstra's algorithm can be found in, \eg,~\cite{PaigeKruskal85} which
gives an EREW PRAM algorithm running in $O(m/p+n\log n)$ parallel time
on $p$ processors. Using a parallel priority queue with constant time
operations, this can be improved to, \eg, $O(n^2/p+n)$ for dense
graphs~\cite{Traff98:ppq}.  The number of sequential steps remain the
bottleneck.

Our aim in the rest of this paper is to be able identify more correct
vertices $u\in F$ at the same time, and do all relaxations from these
vertices in parallel.

\section{Identifying correct vertices}
\label{sec:criteria}

We use a generic SSSP algorithm that runs in \emph{phases} in which
several vertices can be settled in parallel. Let the partition of $V$
into $S$, $F$, and $U$ be as above.  Let $\phi(v)$ be a predicate on
vertices $v\in F$ which we in the following call a
\emph{criterion}. The phased SSSP algorithm initializes $S,F,U$ as
does Dijkstra's algorithm. At the beginning of a phase, a set $S'$
consisting of some, all, but at least one vertex $v\in F$ fulfilling
$\phi(v)$ is computed and removed from $F$. All $v\in S'$ are then
settled in some order, or in parallel, which means that all outgoing,
adjacent edges are relaxed. All vertices of $S'$ are then moved to
$S$, and data structures needed for the next phase updated. The
algorithm terminates when $F=\emptyset$, or when no vertices $v\in F$
fulfill $\phi(v)$.

\begin{definition}[Soundness and completeness]
  A criterion $\psi(v)$ is called \emph{sound} if for each $v\in F$,
  $\psi(v)$ implies $d[v] = \dist(s, v)$, that is $v$ is correct.
  A criterion $\phi(v)$ is called \emph{complete} if whenever
  $F\neq\emptyset$, $\phi(v)$ holds for at least one vertex $v\in F$.
\end{definition}

A sound criterion guarantees that the generic SSSP algorithm computes
only correct distances. A sound and complete criterion ensures that a
shortest path to all reachable vertices from $s$ has indeed been found
(note that our usage of ``sound'' and ``complete'' is
non-standard). Dijkstra's algorithm is the instance of the generic
algorithm with the criterion $\DIJK(v)$ defined by $d[v]=\min_{u\in
  F}d[u]$, where one vertex satisfying $\DIJK(v)$ is selected for $S'$
in each phase.  The correctness proof of Dijkstra's algorithm shows
precisely that $\DIJK(v)$ is sound and complete. It is interesting
that any sound criterion in combination with the settling step will
maintain the invariant on $S,F$, and $U$, with the important property
that edges are relaxed only once. That is, the generic algorithm
maintains the \emph{label setting} property of Dijkstra's algorithm
such that the total work for all relaxations is $O(m)$.  The number of
phases incurred by a specific criterion (on given inputs) is a lower
bound on the time that a parallel algorithm can achieve, regardless of
the number of processors employed.

Let $\phi$ and $\psi$ be two criteria on vertices $v\in F$. We say
that $\psi$ is \emph{stronger} than $\phi$ if
$\phi(v)\Rightarrow\psi(v)$, that is, the stronger criterion
identifies more vertices as correct (note the intentional,
non-standard definition of implied strength). It is then easy to see
that if $\psi$ is sound, then also $\phi$ is sound, and if $\phi$ is
complete, then also $\psi$ is complete. Furthermore, a disjunction of
criteria is sound if all disjuncts are sound, and complete if at least
one of the disjuncts is complete~\cite[Chapter 3]{Kainer18}.

The strongest possible criterion is the clairvoyant \emph{oracle}
criterion $\ORACLE(v)$ defined as $d[v]=\dist(s,v)$ which holds as
soon as $d[v]$ happens to be correct. Clearly, $\DIJK(v)\Rightarrow
\ORACLE(v)$.  The oracle criterion can be used to determine the
smallest number of phases in the generic algorithm and thus the
maximum amount of parallelism for a given input graph, but computing
it efficiently seems to require knowledge of the distance to all
vertices (omniscience, clairvoyance).

We now introduce two new criteria and several weaker variants to be
used in our generic algorithm. The criteria considerably strengthens
the IN and OUT criteria originally proposed by Crauser et
al.~\cite{CrauserMehlhornMeyerSanders98}.

\usetikzlibrary{arrows,calc,decorations.pathmorphing}
\tikzset{snake arrow/.style={
  g->,
  decorate,
  decoration={snake,amplitude=.4mm,segment length=4mm,post length=2mm}
}}
\tikzset{graph node/.style={
  circle,
  draw,
  fill=black,
  inner sep=0pt,
  minimum width=5pt
}}
\tikzset{cost label/.style={
  midway,
  sloped,
  scale=0.5
}}
\tikzset{g->/.style={
  ->,
  >=stealth
}}
\tikzset{g<-/.style={
  g->,
  <-
}}

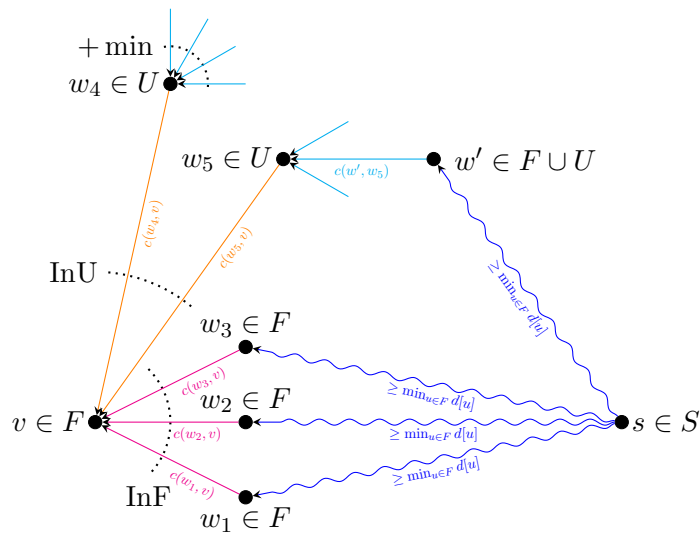
\begin{figure}
  \begin{center}
    \begin{eqnarray*}
      \IN(v) \equiv
      d[v]-\min\left\{
      \begin{array}{c}
        \min_{w\in F, (w,v)\in E}\textcolor{magenta}{c(w,v)} \\
        \min_{w\in U,w'\in F\cup U,(w,v),(w',w)\in E}\textcolor{orange}{c(w,v)}+\textcolor{cyan}{c(w',w)}
      \end{array}
      \right\}
      & \leq &
      \textcolor{blue}{\min_{u\in F}d[u]}
    \end{eqnarray*}
    \begin{tikzpicture}
      \node[graph node]
      (V) at (0,0) {};
      \node[graph node]
      (W1) at (2,-1) {};
      \node[graph node]
      (W2) at (2,0) {};
      \node[graph node]
      (W3) at (2,1) {};

      \node[graph node]
      (W4) at (1,4.5) {};
      \node[graph node]
      (W5) at (2.5,3.5) {};
      
      \node[graph node]
      (S) at (7,0) {};

      \draw (V) node[left] {$v\in F$};
      \draw (S) node[right] {$s\in S$};

      \draw (W1) node[below] {$w_1\in F$};
      \draw (W2) node[above] {$w_2\in F$};
      \draw (W3) node[above] {$w_3\in F$};

      \draw (W4) node[left] {$w_4\in U$};
      \draw (W5) node[above,left] {$w_5\in U$};

      \draw[thick,dotted] ([shift=(85:2)]0,0) arc (85:50:2) node[at start,left] {$\InU$};

      \draw[thick,dotted] ([shift=(45:1)]0,0) arc (45:-45:1) node[at end,below] {$\InF$};

      \draw[thick,dotted] ([shift=(100:.5)]1,4.5) arc (100:-10:.5) node[at start,left] {$+\min$};
      
      \draw[snake arrow,draw=blue] (S) -- node[cost label,below] {$\textcolor{blue}{\geq \min_{u\in F}d[u]}$} (W3);
      \draw[snake arrow,draw=blue] (S) -- node[cost label,below] {$\textcolor{blue}{\geq \min_{u\in F}d[u]}$} (W2);
      \draw[snake arrow,draw=blue] (S) -- node[cost label,below] {$\textcolor{blue}{\geq \min_{u\in F}d[u]}$} (W1);
      \draw[g->,draw=magenta] (W1) -- node[cost label,below,pos=0.3] {$\textcolor{magenta}{c(w_1, v)}$} (V);
      \draw[g->,draw=magenta] (W2) -- node[cost label,below,pos=0.3] {$\textcolor{magenta}{c(w_2, v)}$} (V);
      \draw[g->,draw=magenta] (W3) -- node[cost label,below,pos=0.3] {$\textcolor{magenta}{c(w_3, v)}$} (V);

      \draw[g->,draw=orange] (W4) -- node[cost label,below,pos=0.4] {$\textcolor{orange}{c(w_4, v)}$} (V);
      \draw[g->,draw=orange] (W5) -- node[cost label,below,pos=0.3] {$\textcolor{orange}{c(w_5, v)}$} (V);

      \foreach \a in {0,30,60,90}
      \draw[g<-,draw=cyan] (W4) -- +(\a:1);

      \foreach \a in {-30,0,30} {
        \ifthenelse{\a=0}{
          \draw[g<-,draw=cyan] (W5) -- node[cost label,below]{$\textcolor{cyan}{c(w', w_5)}$} +(\a:2);
        }{
          \draw[g<-,draw=cyan] (W5) -- +(\a:1);
        }
      }

      \node[graph node]
      (WW) at (4.5,3.5) {};

      \draw (WW) node[right,xshift=4] {$w'\in F\cup U$};
      \draw[snake arrow,draw=blue] (S) -- node[cost label,below] {$\textcolor{blue}{\geq \min_{u\in F}d[u]}$} (WW);      
    \end{tikzpicture}
  \end{center}
  \caption{Structure of possible shortest paths to vertex $v\in F$,
    showing why the $\IN(v)$ criterion correctly decides if $v$ is
    already correct, $d[v]=\dist(s,v)$. Vertex $v$ is correct if
    $d[v]$ is smaller than or equal to the length of each of the shown
    possible paths from $s$. The minima $\InF$ over all
    incoming edges from vertices $w_1,w_2,w_3,\ldots\in F$ and $\InU$
    over all possible paths of two edges from vertices
    $w_4,w_5,\ldots\in U$ that need to be maintained efficiently when
    implementing the criterion are also illustrated. The crucial
    observation for the $\IN(v)$ criterion is that all edges ending in
    $w_4,w_5,\ldots\in U$ must start from vertices $w'\in F\cup U$; if
    not, $w_4,w_5,\ldots$ cannot be in $U$.}
  \label{fig:incriteria}
\end{figure}

\begin{figure}
  \begin{center}
    \begin{eqnarray*}
      \OUT(v) \equiv
      d[v] & \leq &
      \min\left\{
      \begin{array}{c}
        \min_{u\in F,w\in F,(u,w)\in E} \textcolor{gray}{d[u]}+\textcolor{blue}{c(u,w)} \\
        \min_{u\in F, w\in U,w'\in F\cup U,(u,w),(w,w')\in E} \textcolor{gray}{d[u]}+\textcolor{orange}{c(u,w)}+\textcolor{magenta}{c(w,w')}
      \end{array}
      \right\}
    \end{eqnarray*}
    \begin{tikzpicture}[scale=0.90]
      \node[graph node]
      (V) at (0,0) {};
      \node[graph node]
      (U1) at (5,0) {};
      \node[graph node]
      (U2) at (7,1) {};

      \node[graph node]
      (W1) at (4,2) {};

      \node[graph node]
      (W2) at (6,4) {};
      \node[graph node]
      (W3) at (8,5) {};

      \node[graph node]
      (WW4) at (4,4) {};

      \node[graph node]
      (S) at (6.5,-2) {};

      \draw (V) node[left] {$v\in F$};
      \draw (U1) node[right] {$u_1\in F$};
      \draw (U2) node[right] {$u_2\in F$};
      \draw (W1) node[right] {$w_1\in U$};
      \draw (W2) node[right] {$w_2\in U$};
      \draw (W3) node[right] {$w_3\in U$};
      \draw (S) node[right] {$s\in S$};

      \draw[g->,draw=orange] (U1) -- node[cost label,above] {$\textcolor{orange}{c(u_1,w_1)}$} (W1);
      \draw[g->,draw=orange] (U2) -- node[cost label,above,pos=0.4] {$\textcolor{orange}{c(u_2,w_2)}$} (W2);
      \draw[g->,draw=orange] (U2) -- node[cost label,above,pos=0.35,rotate=180] {$\textcolor{orange}{c(u_2,w_3)}$} (W3);
      
      \foreach \a in {150,180,210} {
        \node[graph node] (WW\a) at ([shift={(\a:2)}]U1) {};
        \draw[g->,draw=blue] (U1) -- node[cost label,above,pos=0.77] {
          \ifthenelse{\a=210}{$\textcolor{blue}{c(u_1,w_4)}$}{}
        } (WW\a);
        \draw[snake arrow] (WW\a) -- node[cost label,below] {$\geq 0$} (V);
      }

      \draw (WW210) node[below] {$w_4\in F$};

      \foreach \a in {165,195}
      \draw[g->,color=blue] (U2) -- +(\a:1);

      \foreach \a in {165,195}
      \draw[g->,draw=magenta] (W1) -- +(\a:1);

      \foreach \a in {150,120} {
        \draw[g->,draw=magenta] (W2) -- +(\a:1) node (UUU\a) {};
      }
      \draw[g->,draw=magenta] (W2) -- node[cost label,below] {$\textcolor{magenta}{c(w_2,w')}$} (WW4);

      \foreach \a in {150,120,90,60}
      \draw[g->,>=stealth,draw=magenta] (W3) -- +(\a:1);

      \draw[thick,dotted] ([shift=(135:1)]5,0) arc (135:225:1) node[at end,below,xshift=5pt] {$\OutF$};

      \draw[thick,dotted] ([shift=(120:2)]7,1) arc (120:62:2) node[at end,right] {$\OutU$};

      \draw[thick,dotted] ([shift=(165:.5)]8,5) arc (165:45:.5) node[at end,right] {$+\min$};

      \draw[snake arrow] (WW4) node[above,left] {$w'\in F\cup U$} -- node[cost label,below] {$\geq 0$} (V);

      \draw[snake arrow,draw=gray] (S) -- node[cost label,above] {$\textcolor{gray}{\geq \min_{u\in F}d[u]}$} (U1);

      \draw[snake arrow,draw=gray] (S) -- node[cost label,above,rotate=180] {$\textcolor{gray}{\geq \min_{u\in F}d[u]}$} (U2);
    \end{tikzpicture}
  \end{center}
  \caption{Structure of possible shortest paths to vertex $v\in F$,
    showing why the $\OUT(v)$ criterion correctly decides if $v$ is
    already correct, $d[v]=\dist(s,v)$. Vertex $v$ is correct if
    $d[v]$ is smaller than or equal to each of the possible shortest
    paths from vertices $u_1,u_2,\ldots\in F$ shown. The minima
    $\OutF$ and $\OutU$ that need to be maintained efficiently when
    implementing the criterion are also illustrated. Note that the
    minima over the outgoing edges of vertices $w_1,w_2,w_3,\ldots\in U$
    for the $\OUT(v)$ criterion are over edges with endpoint in $F\cup
    U$, which is costly to maintain efficiently since they change as
    vertices adjacent to $w_1,w_2,w_3,\ldots$ become settled. The
    $\OUTWEAK(v)$ criterion therefore uses only statically computed
    minima over all adjacent edges.}
  \label{fig:outcriteria}
\end{figure}
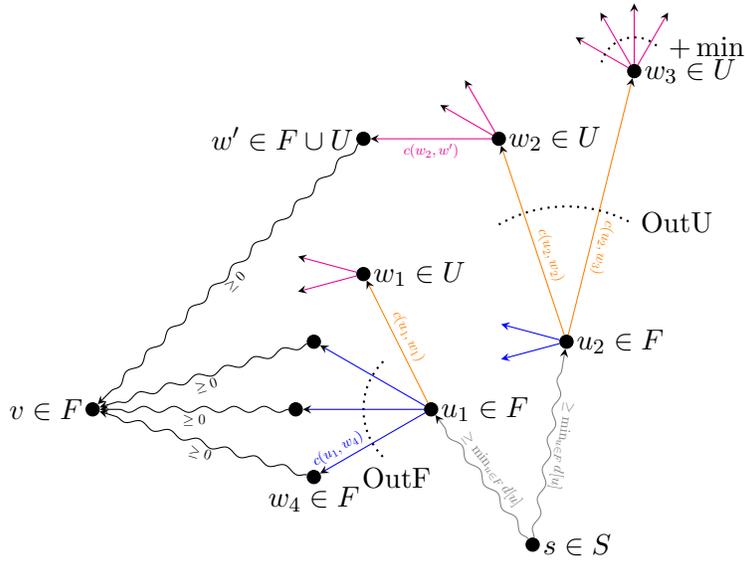

Define the $\IN(v)$ criterion to hold if
\begin{eqnarray}
  d[v]-\min\left\{
  \begin{array}{c}
    \min_{w\in F, (w,v)\in E}c(w,v) \\
    \min_{w\in U,w'\in F\cup U,(w,v),(w',w)\in E}c(w,v)+c(w',w)
  \end{array}
  \right\}
  & \leq &
  \min_{u\in F}d[u]
  \label{crit:in}
\end{eqnarray}

Define the $\OUT(v)$ criterion to hold if
\begin{eqnarray}
  d[v] & \leq &
  \min\left\{
  \begin{array}{c}
    \min_{u\in F,w\in F,(u,w)\in E} d[u]+c(u,w) \\
    \min_{u\in F, w\in U,w'\in F\cup U,(u,w),(w,w')\in E} d[u]+c(u,w)+c(w,w')
  \end{array}
  \right\}
  \label{crit:out}
\end{eqnarray}

\begin{lemma}
  \label{lem:in}
  The $\IN(v)$ criterion defined by Equation~\eqref{crit:in} is sound
  and complete.
\end{lemma}

\begin{proof}
  We have to prove that whenever $\IN(v)$ holds, then
  $d[v]=\dist(s,v)$ such that $v$ is correct.  First observe that for
  any incorrect vertex $v\in F$ with $d[v]>\dist(s,v)$, there is a
  shortest path to $v$ consisting of first a path over vertices in $S$, an
  edge between a vertex in $S$ and a vertex in $F$, followed by a path
  of length at least one over vertices in $F\cup U$. Thus, a shortest
  path to $v$ must end with an edge $(w,v)$ with $w\in F\cup U$. Now
  assume that $v\in F$ is not correct, but that for all $w\in F$,
  $d[v]-c(w,v)\leq \min_{u\in F}d[u]$, that is $\IN(v)$ holds. Since
  $\min_{u\in F}d[u]\leq\dist(s,w)$, it follows that
  $d[v]\leq\dist(s,w)+c(w,v)$ for all $w\in F$, especially the $w$ for
  which $\dist(s,v)=\dist(s,w)+c(w,v)$, contradicting that
  $d[v]>\dist(s,v)$. A shortest path to $v$ for which the last edge is
  some $(v,w)\in E$ with $w\in U$ must have at least two edges outside
  of $S$, since for any $w\in U$ there cannot be an edge $(w',w)$ with
  $w'\in S$ (due to the relaxation when a vertex is settled and moved
  to $S$). Also $d[v]-(c(w',w)+c(w,v))<\min_{u\in F}d[u]$ for all such
  two edges would contradict that $d[v]$ is not correct. Note that
  $w'\in F\cup U$, since $w'\in S$ would contradict that $w\in U$ has
  not been explored.

  For completeness, observe that any vertex $v$ with $d[v]=\min_{u\in
    F}d[u]$ fulfills $\IN(v)$.
\end{proof}
  
\begin{lemma}
  \label{lem:out}
  The $\OUT(v)$ criterion defined by Equation~\eqref{crit:out} is sound
  and complete.
\end{lemma}

\begin{proof}
A shortest path to $v\in F$ has a first vertex $u$ which is not in
$S$. This vertex must be in $F$, and $d[u]$ must be the correct
distance to this vertex, $d[u]=\dist(s,u)$.  Assume that $v$ is not
correct, but that $\OUT(v)$ holds in which case there is at least one
edge $(u,w),w\neq v$ on a shortest path to $v$.  Then $d[v]\leq
d[u]+c(u,w)\leq\dist(s,v)$ for $w\in F$ contradicts that
$\dist(s,v)<d[v]$. If $w$ is in $U$, a shortest path to $v$ must have
at least one more edge $(w,w')\in E$ with $w'\in F\cup U$, and
$d[v]\leq d[u]+(c(u,w)+c(w,w'))\leq\dist(s,v)$ again contradicts the
assumption that $d[v]>\dist(s,v)$.

For completeness, for any vertex $u$ leading to the minimum value at the
right hand side of Equation~\eqref{crit:out}, $\OUT(u)$ will hold.
\end{proof}

The two criteria are orthogonal. There are vertices $v$ that are
correct according to the $\IN(v)$ criterion, but not according to the
$\OUT(v)$ criterion, and vice versa. The two criteria can be combined
disjunctively to further reduce the number of phases.  Note that the
completeness argument for the $\IN(v)$ criterion shows that
$\DIJK(v)\Rightarrow\IN(v)$. This is not the case for the $\OUT(v)$
criterion which may choose a different vertex than one having minimum
tentative distance. The structure of the paths establishing the
$\IN(v)$ and $\OUT(v)$ criteria (and the weaker variants discussed in
the following) are illustrated in Figures~\ref{fig:incriteria}
and~\ref{fig:outcriteria}. These figures also illustrate the minima
maintained with the data structures described in the algorithms
discussed in the following propositions.

We now claim that we can implement the generic SSSP algorithm using
the $\IN(v)$ criterion in $O(m\log n)$ operations.

\begin{proposition}
  \label{prop:inphase}
  The generic SSSP algorithm exploiting the $\IN(v)$ criterion of
  Equation~\eqref{crit:in} can be implemented to run in
  $O(n\log n+m\log n)$ operations.
\end{proposition}

For the implementation, we assume that the input graph is given as an
array of adjacency lists of both outgoing and incoming edges for each
vertex. Additionally, for each incoming edge $(u,v)$ of $v$, there is
a reference to the position of $(u,v)$ in the list of outgoing edges
of $u$. Similarly for the outgoing edges. If this is not the case,
such a representation can be computed in $O(n+m)$ operations.

\begin{proof}
  As for Dijkstra's algorithm, we use a priority queue of tentative
  distances $d[v]$ for $v\in F$ supporting \Qdelmin, \Qinsert, and
  \Qdecrease operations.
  
  We associate two heaps (priority queues) $\InF[v], \InU[v]$
  supporting \Qfindmin, \Qdelete, \Qinsert and \Qbuild operations with
  each vertex $v\in F\cup U$. These heaps will store edge costs
  corresponding to the two $\min$-terms in the left hand side of the
  IN criterion, such that the left hand side expression can be
  computed as $d[v]-\min(\Qfindmin\,\InF[v],\Qfindmin\,\InU[v])$.  With
  the \Qfindmin heap operation taking constant time, it can be
  determined in constant time for any $v\in F$ whether the IN
  criterion is satisfied. Keeping the values $d[v]-\min(\Qfindmin\,%
  \InF[v],\Qfindmin\,\InU[v])$ in yet another priority queue $\InCrit$,
  the vertices $v$ for which the $\IN(v)$ criterion are satisfied can be
  extracted in $O(\log n)$ operations each.

  The generic algorithm is instantiated as follows. Initially, for
  each $v\in U$, for all incoming edges, the costs $(\min_{w'\in F\cup
    U, w\in U, (w',w)\in E}c(w',w))+c(w,v)$ are inserted into
  $\InU[v]$ by a bulk build heap operation. The crucial observation
  here is that for $w\in U$, it will always hold that for all incoming
  edges $(w',w)\in E$ that $w'\in F\cup U$ (if there were a vertex
  $w'\in S$, the edge $(w',w)$ would have been relaxed, and $w$ not in
  $U$). The value $(\min_{w'\in F\cup U, w\in U, (w',w)\in E}c(w',w))$
  will therefore not change as long as $w\in U$ and can be precomputed
  for all vertices in $O(m)$ operations.  In other words, the term
  $\min_{w\in U,w'\in F\cup U,(w,v),(w',w)\in E}c(w,v)+c(w',w)$ is
  computed as $\min_{w\in U,(w,v)\in E}(c(w,v)+\min_{w'\in V,(w',w)\in
    E}c(w',w))$ The other heap $\InF[v]$ shall be empty for all $v\in
  V$.

  Let $v\in F$ be a vertex satisfying $\IN(v)$.  When vertex $v$ is
  settled and moved to $S$, all outgoing edges $(v,w)\in E$ with $w\in
  F\cup U$ are relaxed. If $w\in F$, the edge cost $c(v,w)$ is deleted
  from $\InF[w]$. If $w\in U$ and therefore visited for the first time
  and moved to $F$, all outgoing edges $(w,w')$ with $w'\in F\cup U$
  have to be scanned in order to maintain the invariants on the
  heaps. If $w'\in U$, the cost $(\min_{u\in F\cup U, w\in U,
    (w',w)\in E}c(u,w))+c(w,w')$ is deleted from $\InU[w']$. If $w'\in
  F$, the cost $c(w,w')$ is inserted into the heap $\InF[w']$.  Since
  $v$ is moved to $F$ from $U$ only once, there are at most $m$ such
  heap operations in total. With heap insert and delete operations
  taking $O(\log n)$ operations, the total number of operations is
  $O(m\log n)$ as claimed.

  Also the $\InCrit$ priority queue has to be updated. This is done by
  keeping track of all vertices $v$ for which either $d[v]$,
  $\mathrm{InU}[v]$ or $\mathrm{InF}[v]$ change in a phase. At the end
  of the phase, the value in InCrit for these vertices is decreased
  accordingly. In total, there are at most $m$ such changes over the
  execution of the algorithm. 
\end{proof}

For the $\OUT(v)$ criterion, the second minimum term in the right hand
side of Equation~\eqref{crit:out} does change throughout the execution
of the algorithm. Since the minimum is over edges $(w,w')$ with $w'\in
F\cup U$, and $F$ is updated each time a vertex is settled, the minimum
can increase. Therefore an $\OutU[w]$ heap to decide whether the
second minimum term of $\OUT(v)$ is satisfied cannot be maintained as
for the $\IN(v)$ criterion. It does not seem possible to implement the
$\OUT(v)$ criterion in $O(m\log n)$ operations. However, if we weaken
the criterion slightly, the same ideas as in
Proposition~\ref{prop:inphase} can be employed.

Define the $\OUTWEAK(v)$ criterion to hold if
\begin{eqnarray}
  d[v] & \leq &
  \min\left\{
  \begin{array}{c}
    \min_{u\in F,w\in F,(u,w)\in E} d[u]+c(u,w) \\
    \min_{u\in F, w\in U,w'\in V,(u,w),(w,w')\in E} d[u]+c(u,w)+c(w,w')
        \end{array}
  \right\}
  \label{crit:outweak}
\end{eqnarray}

The difference to the stronger $\OUT(v)$ criterion, is that the
$\min_{u\in F, w\in U,w'\in F\cup U,(u,w),(w,w')\in E}
d[u]+c(u,w)+c(w,w')$ is approximated by the possibly smaller
$\min_{u\in F, w\in U,(u,w)\in E} d[u]+(c(u,w)+\min_{w'\in V,(w,w')\in
  E}c(w,w'))$, such that $\OUTWEAK(v)\Rightarrow\OUT(v)$

\begin{proposition}
  \label{prop:outphase}
  The generic SSSP algorithm exploiting the $\OUTWEAK(v)$ criterion of
  Equation~\eqref{crit:outweak} can be implemented to run in
  $O(n\log n+m\log n)$ operations.
\end{proposition}

\begin{proof}
 As for Dijkstra's algorithm, we use a priority queue of tentative
 distances $d[v]$ for $v\in F$ supporting \Qdelmin, \Qinsert, and
 \Qdecrease operations.

  We associate two heaps (priority queues) $\OutF[v], \OutU[v]$
  supporting \Qfindmin, \Qdelete, \Qinsert and \Qbuild operations with
  each vertex $v\in F\cup U$. The $\OutF[v]$ heap stores the costs
  $c(v,w)$ for $w\in F,(v,w)\in E$, and the $\OutU[v]$ heap the costs
  $c(v,w)+\min_{w\in U,w'\in V, (w,w')\in E}c(w,w')$. The right hand
  side of Equation~\eqref{crit:outweak} can be computed as $d[u]$ plus
  the minimum of the two heaps, and we keep this for all $u$ in a
  third priority queue $\OutCrit$. At the beginning of a phase, all
  vertices in the priority queue of tentative distances that are
  smaller than the minimum value in $\OutCrit$ will fulfill
  $\OUTWEAK(v)$. Values are deleted from $\OutCrit$ as the
  corresponding vertices become settled.

  When a vertex $u$ is settled, the cost $c(u,w)$ is deleted from the
  heap $\OutF[w]$ for all $w\in F$. If $w\in U$ meaning that $w$ moves
  to $F$, the incoming edges $(w',w)$ with $w'\in F\cup U$ are
  scanned. If $w'\in F$, the cost $c(w',w)$ inserted into
  $\OutF[w']$. If $w'\in U$, the cost $c(w',w)+\min_{v\in V,(w,v)\in
    E}c(w,v)$ is deleted from $\OutU[w']$. Since a vertex moves to $F$
  once at most, the total number of heap operations is at most $m$. At
  the end of the phase, the values in $\OutCrit$ are increased for all
  $v$ for which the minimum values in $\OutF[v]$ or $\OutU[v]$ have
  changed. When a $d[u]$ value change in some relaxation step, the
  value in $\OutCrit$ is decreased. The $\OutCrit$ priority queue must
  therefore support both \Qdecrease and \Qincrease operations
  efficiently.
\end{proof}

The heaps associated with the vertices can all be eliminated, and the
implementation considerably simplified by first presorting the edges in
increasing cost order.
\begin{proposition}
  \label{prop:presorted}
  With a presorting of the incoming and outgoing edges of all vertices
  in order of increasing cost taking $O(n+m\log n)$ operations, the
  generic SSSP algorithm exploiting the $\IN(v)$ and $\OUTWEAK(v)$
  criteria can be implemented in $O(n\log n+m)$ operations.
\end{proposition}

\begin{proof}
  Four presorting steps are needed, and edge lists need to be
  maintained as doubly linked lists to support easy deletion of
  incoming and outgoing edges. For each vertex $u\in V$, a list of
  incoming edges in increasing cost order, and a list of outgoing
  edges in increasing cost order is constructed. For each $v$, the
  minima $M[v]=\min_{(v,w)\in E}c(v,w)$ and $M'[v]\min_{(u,v)\in
    E}c(u,v)$ are precomputed and sorted lists of $c(v,w)+M[w]$ and
  $M'[u]+c(u,v)$ are constructed. Priority queues $\InCrit$ and
  $\OutCrit$ are used in addition to a priority queue of tentative
  distances, and maintained as in the algorithms of
  Proposition~\ref{prop:inphase} and
  Proposition~\ref{prop:outphase}. With these, the vertices for which
  the $\IN(v)$ and $\OUTWEAK(v)$ criteria hold can be selected in
  $O(\log n)$ time each.

  The minima in the right hand sides of the criteria can be found in
  constant time per vertex, simply by looking at the first element in
  the corresponding sorted adjacency list. When a vertex $u$ is
  settled, the edges $(v,u)$ are removed from the lists of outgoing
  edges of all $v\in F$.  Also, when the edges $(u,v)$ are relaxed,
  these edges are removed from the lists of incoming edges for all
  $v$. When in the relaxation some vertex $v\in U$ is seen for the
  first time and moved to $F$, the remaining outgoing edges $(v,w)$
  are scanned, and the cost $c(w,v)+M[v]$ removed from the list of
  these costs of $w$. Also the cost $M'[w]+c(v,w)$ is removed from the
  list of these costs of $v$.
  
  When the removal of an edge from one of the edge lists causes a
  minimum to change, the corresponding vertex is marked, and at the
  end of the phase the values in the $\InCrit$ and $\OutCrit$ priority
  queues are adjusted accordingly.
\end{proof}

Weaker, but possibly easier to compute criteria can be derived from
these two criteria by taking minima over larger sets and/or over
smaller values. All in all, this will lead to smaller minima in the
criteria, and thus weaker, because more restrictive, criteria. Such
considerations give rise to the following derived criteria.

The weakest criteria we consider use minima over all edges. We
call these criteria \emph{static} because the minima can be computed
in advance and are not changed during the execution of the generic
algorithm. The static criteria are the criteria originally introduced
by Crauser et al.~\cite{CrauserMehlhornMeyerSanders98}.

Define $\INSTATIC(v)$ to hold if
\begin{eqnarray}
  d[v]-\min_{w\in V, (w,v)\in E}c(w,v) & \leq & \min_{u\in F}d[u]
  \label{crit:instatic}
\end{eqnarray}

Define $\OUTSTATIC(v)$ to hold if
\begin{eqnarray}
  d[v] & \leq & \min_{u\in F,w\in V,(u,w)\in E} d[u]+c(u,w)
  \label{crit:outstatic}
\end{eqnarray}

Since the minima in both cases do not change during the execution of
the algorithm, and can therefore be precomputed in $O(m)$ operations,
the generic, phased SSSP algorithm can be implemented sequentially in
$O(n\log n+m)$ operations with the $\INSTATIC(v)$ and $\OUTSTATIC(v)$
criteria.

It might be expected that the minimum terms concerning the unexplored
vertices $U$ in both $\IN(v)$ and $\OUT(v)$ criteria will not bring
much, since graphs with any expansion properties will quickly lead to
$U=\emptyset$ in the algorithm. For this reason, criteria without
these terms are considered as given in Equation~\eqref{crit:insimple}
and Equation~\eqref{crit:outsimple}.  The fact that any predecessor,
respectively successor, in $U$ actually enforces at least two vertices
not in $S$ to be present in a shortest path to $v$ is not exploited in
these criteria. The $U$ case is simply subsumed under the $F$ case
which considers only a single edge on a shortest path and therefore
leads to a potentially smaller minimum. The main advantage of these
criteria is again a potentially more efficient implementation as was
the case with the $\OUTWEAK(v)$ criterion. We call these weakenings
the \emph{simple, dynamic} criteria. The criteria are called dynamic,
since the minima do change over the course of the execution, and thus
gradually strengthen of the criteria.

Define $\INSIMPLE(v)$ to hold if
\begin{eqnarray}
  d[v]-\min_{w\in F\cup U, (w,v)\in E}c(w,v) & \leq & \min_{u\in F}d[u]
  \label{crit:insimple}
\end{eqnarray}

Define $\OUTSIMPLE(v)$ to hold if
\begin{eqnarray}
  d[v] & \leq & \min_{u\in F\cup U,w\in V,(u,w)\in E} d[u]+c(u,w)
  \label{crit:outsimple}
\end{eqnarray}

We have argued that
$\DIJK(v)\Rightarrow\INSTATIC(v)\Rightarrow\INSIMPLE(v)\Rightarrow\IN(v)$,
and likewise
$\OUTSTATIC(v)\Rightarrow\OUTSIMPLE(v)\Rightarrow\OUTWEAK(v)\Rightarrow\OUT(v)$.

Finally, we discuss yet another way of instantiating the generic SSSP
algorithm which may be attractive to implement. The idea here is to
approximate the criteria by recomputing the minima at certain
intervals, under certain conditions.  We use the same priority queues
and basic idea as in the implementation in
Proposition~\ref{prop:presorted}, with doubly linked lists, but do not
presort any edge lists. Edges are eliminated as vertices move to $S$
and to $F$, but now exact minima cannot be looked up in constant
time. Instead, with each vertex, four approximate minima shall be
maintained, namely $\min_{w\in F}c(w,v)$ and $\min_{w\in
  U}(c(w,v)+\min_{w'\in V}c(w',w))$ for the $\IN(v)$ criterion, and
$\min_{w\in F}c(v,w)$ and $\min_{w\in U}(c(v,w)+\min_{w'\in
  V}c(w,w'))$ for the $\OUTWEAK(v)$ criterion.

The minima are maintained conservatively, and might be too small, but
this does not invalidate the decisions made by the criteria.  During
the execution of the algorithm, when either of these minima might
change, due to a vertex moving to either $S$ for $F$, the
corresponding minimum is recomputed. A potential change to a minimum
can be detected by also keeping track of the incoming or outgoing edge
corresponding to the minimum value. Recomputation, on the other hand,
cannot be afforded at every change (of which there are $O(m)$), so a
parameter $k$ is chosen, and minima recalculated at most at every
$k$th change.

The resulting, approximate criteria are not comparable to the stricter
ones.  Since they may not be complete, they need to be combined with
$\DIJK(v)$. A particular instance of the approximation is when minima
are computed only once as in~\cite{Garg18}. This choice leads to
criteria at least as strong as $\INSTATIC(v)$ and $\OUTSTATIC(v)$.

\section{Simulations}
\label{sec:simulations}

In this section we explore the reduction in the number of phases that can be
achieved by applying and combining the criteria introduced in the
previous section.%
\footnote{%
The source code of the simulation tool can be downloaded at
\url{https://github.com/kaini/sssp-simulation}.%
}
We investigate the following combinations of criteria:

\begin{itemize}
\item
  The strongest criteria $\IN(v)$, $\OUT(v)$, and the disjunction
  $\IN(v)\vee\OUT(v)$. These are referred to as the full criteria in the plots.
\item
  The simple, dynamic criteria $\INSIMPLE(v)$, $\OUTSIMPLE(v)$, and
  the disjunction $\INSIMPLE(v)\vee\OUTSIMPLE(v)$
\item
  The original, static criteria by Crauser et al., $\INSTATIC(v)$,
  $\OUTSTATIC(v)$, and the disjunction $\INSTATIC(v)\vee\OUTSTATIC(v)$
\item
  The oracle criterion $\ORACLE(v)$
\end{itemize}

We measure the number of phases in a generic SSSP algorithm where all
vertices fulfilling the criteria prior to a phase are selected and
settled. We also estimate the amount of work needed to find vertices
in the fringe set $F$ by summing the sizes $|F|$ over all phases. This
provides information on the data structure support required for a real,
efficient implementation of the generic algorithm.

The first set of simulations was performed on \emph{uniformly random
  graphs} which is the same family of graphs used in the paper by
Crauser et al.. A uniformly random graph $G(n, p)$ consists of $n$
vertices where the~(independent) probability for each edge is
$p$. Since there are $n(n-1)$ possible edges the number of edges in
the graph is distributed as $\mathrm{binom}(n(n-1), p)$, which has an
expected value of $n(n-1)p$, or approximately $n^2p$. The edge weights
are uniformly distributed in the range~$[0; 1]$.

The second set of simulations was performed on \emph{Kronecker
  graphs}~\cite{Leskovec10}. Kronecker graphs are generated by
repeatedly multiplying a small, so-called (square) initiator matrix of
positive real numbers with itself by utilizing the Kronecker
product\footnote{The Kronecker product $\otimes$ is defined as
  $ A \otimes B =
    \begin{pmatrix}
        a_{1,1}B   & \dots  & a_{1,n_A}B   \\
        \vdots     & \ddots & \vdots       \\
        a_{m_A,1}B & \dots  & a_{m_A,n_A}B
    \end{pmatrix}
    $. }. The result represents the probability for each single
possible edge to appear in the sampled graph. The edges are
unweighted.  For the purposes of these simulations the initiator
matrix is $2.5\left(\begin{array}{cc}0.57 & 0.19\\ 0.19 &
  0.05\end{array}\right)$ as also used in the Graph~500
benchmark~\cite{graph500}. The multiplication with~$2.5$ is to
control the number of edges of the final graph. In our implementation
for sampling Kronecker graphs, the expected
number of edges in the resulting graph is
$(\sum \mathrm{initiator~matrix})^k$ with~$k$ being
the Kronecker exponent. If we would use the initiator matrix as it is,
this would be~1, \ie, on average a single edge would be generated,
no matter the value of~$k$. Of course,
the multiplication does not change the structure of the graph in any way,
it is just a way to control the number of generated edges.
Leskovec et al.\ claim that
Kronecker graphs have many properties that are present in real-world
networks, like social networks or citation networks. Furthermore,
they provide a fast algorithm to generate such graphs.

\begin{figure}
    \centering
    \input{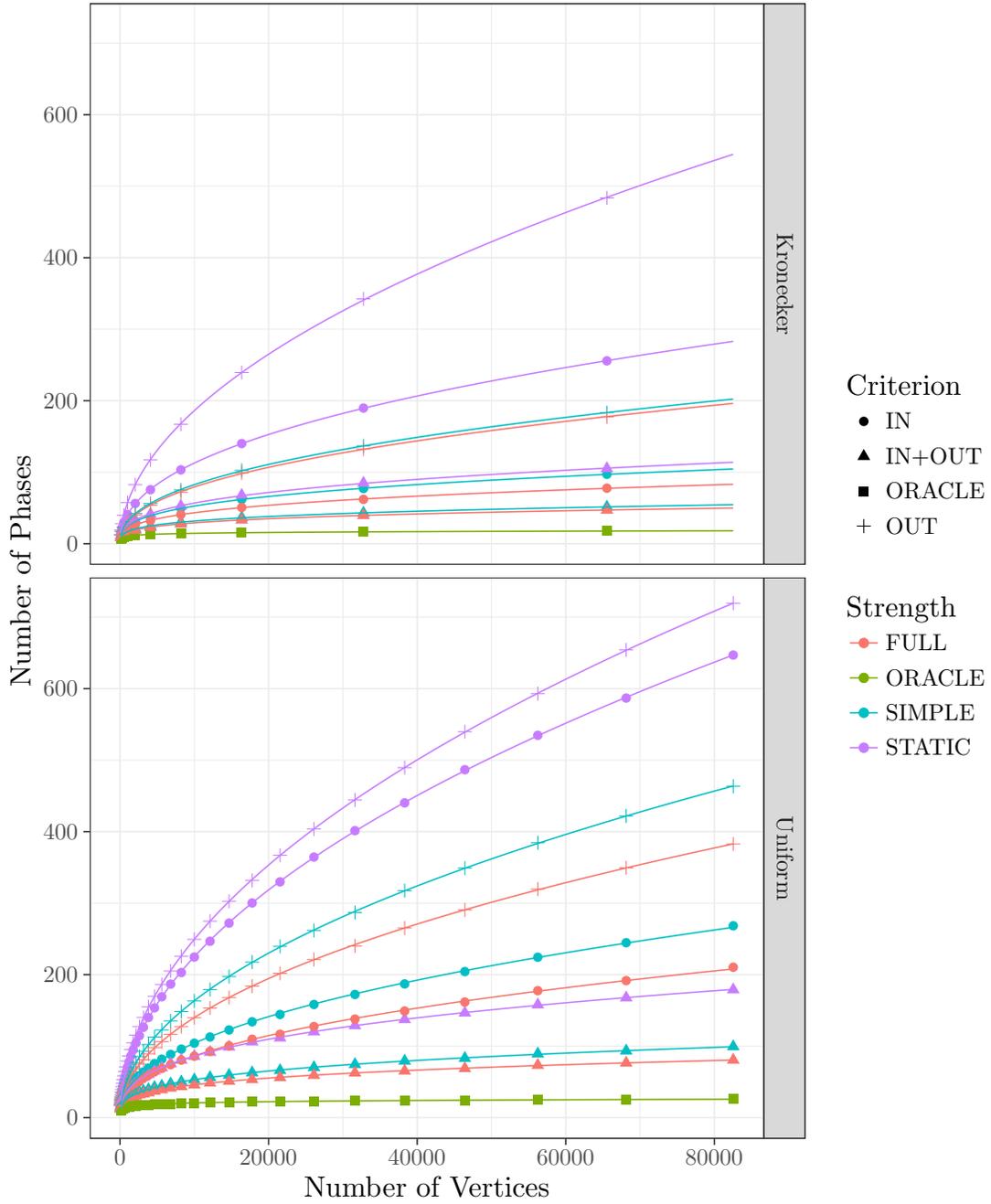}
    \caption{The number of phases required by the criteria on uniform
      graphs with an expected out-degree of~10 for each vertex and
      Kronecker graphs with the initiator matrix
      $2.5\left(0.57 \ 0.19; 0.19 \ 0.05\right)$.}
    \label{simulation_phases}
\end{figure}

Figure~\ref{simulation_phases} shows the mean number of phases
required to settle all reachable vertices of a sample of~100 graphs
for the given class. The uniform graphs~$G(n, p)$ are generated in
such a way that $m/n = 10$, \ie, care is taken that the expected
out-degree of vertices of these graphs stays constant when increasing
the number of vertices. The used classes are therefore $G(100,
0.1010)$, $G(121, 0.0833)$, $G(147, 0.0685)$, and so on. The Kronecker
graphs were generated by sampling edges from
$\left(2.5\left(\begin{array}{cc}0.57 & 0.19\\ 0.19 &
  0.05\end{array}\right)\right)^k$ ranging from~$k=7$ to~$k=16$,
\ie, vertex counts
from~128~($2^7$) to~65536~($2^{16}$).  Edge weights are uniformly
distributed in~$[0;1]$.

For uniform graphs, $\INSTATIC(v)\vee\OUTSTATIC(v)$, \ie, the weakest
disjunctive criterion we discuss, already beats all non-disjunctive
criteria. This means that it is possible to achieve a reasonably small
number of phases with a quite simple criterion and low implementation
complexity.  $\INSIMPLE(v)\vee\OUTSIMPLE(v)$ and to a lesser extend
$\IN(v)\vee\OUT(v)$ further improves on the number of phases by a
factor of about~$1.8$, respectively $1.2$. Nevertheless, $\ORACLE(v)$
is still unreached by all criteria.  Compared to $\IN(v)\vee\OUT(v)$
the oracle only needs a third of the number of phases.  For Kronecker
graphs the results are similar, except that $\INSIMPLE(v)$ and
$\IN(v)$ are stronger than $\INSTATIC(v)\vee\OUTSTATIC(v)$. Again, the
$\ORACLE(v)$ criterion is unreached.

We performed curve-fitting to obtain numerical estimations of the
number of phases for the various criteria. These can be seen in
Table~\ref{phases_curve_fitting}. Each result was fitted using the
functions $a + b \cdot \log_2(n)$ and $a + b \cdot n^c$ with the
parameters $a$, $b$, and $c$. The most appropriate fit was chosen for
the data in Table~\ref{phases_curve_fitting}.  For simplicity the
parameter~$a$ was dropped from the results.  A notable result is that
for uniform graphs $\INSTATIC(v)$, $\OUTSTATIC(v)$, $\INSIMPLE(v)$,
$\OUTSIMPLE(v)$, $\IN(v)$, and $\OUT(v)$ have almost the same
exponent.  The various criteria only change the multiplicative
factor. Only the disjunctive criteria reduce the exponent from
about~$1/2$ down to~$1/3$ and~$1/4$.  $\ORACLE(v)$ only needs a
logarithmic number of phases. In other words there is still a
considerable gap between our criteria and the oracle.  For Kronecker
graphs it seems that the family of $\IN$-criteria is stronger than the
corresponding family of $\OUT$-criteria, additionally the simple
criteria have a higher influence on the exponent than is the case for
uniform graphs. Nevertheless, the structure of the results is very
similar to uniform graphs.

\begin{table}[]
  \centering
  \caption{The number of phases required by various criteria to settle
    all vertices in uniform and Kronecker graphs. The numbers were
    obtained by curve-fitting.}
  \label{phases_curve_fitting}
  \begin{tabular}{r|c|c}
    \textbf{Criterion} & \textbf{Uniform Graphs} & \textbf{Kronecker Graphs} \\
    \hline
    $\OUTSTATIC(v)$ & $2.48 \cdot n^{0.5}$ & $1.79 \cdot n^{0.51}$ \\
    $\INSTATIC(v)$ & $2.28 \cdot n^{0.5}$ & $2.17 \cdot n^{0.43}$ \\
    $\OUTSTATIC(v)\vee\INSTATIC(v)$ & $3.97 \cdot n^{0.34}$ & $3.49 \cdot n^{0.31}$ \\
    \hline
    $\OUTSIMPLE(v)$ & $1.66 \cdot n^{0.5}$ & $1.68 \cdot n^{0.42}$ \\
    $\INSIMPLE(v)$ & $1.43 \cdot n^{0.46}$ & $3.01 \cdot n^{0.32}$ \\
    $\OUTSIMPLE(v)\vee\INSIMPLE(v)$ & $3.75 \cdot n^{0.29}$ & $4.03 \cdot n^{0.24}$ \\
    \hline
    $\OUT(v)$ & $1.62 \cdot n^{0.48}$ & $1.54 \cdot n^{0.43}$ \\
    $\IN(v)$ & $1.47 \cdot n^{0.43}$ & $2.83 \cdot n^{0.3}$ \\
    $\OUT(v)\vee\IN(v)$ & $4.60 \cdot n^{0.26}$ & $3.65 \cdot n^{0.24}$ \\
    \hline
    $\ORACLE(v)$ & $1.69 \cdot \log_2(n)$ & $1.17 \cdot \log_2(n)$
  \end{tabular}
\end{table}

\begin{figure}
    \centering
    \input{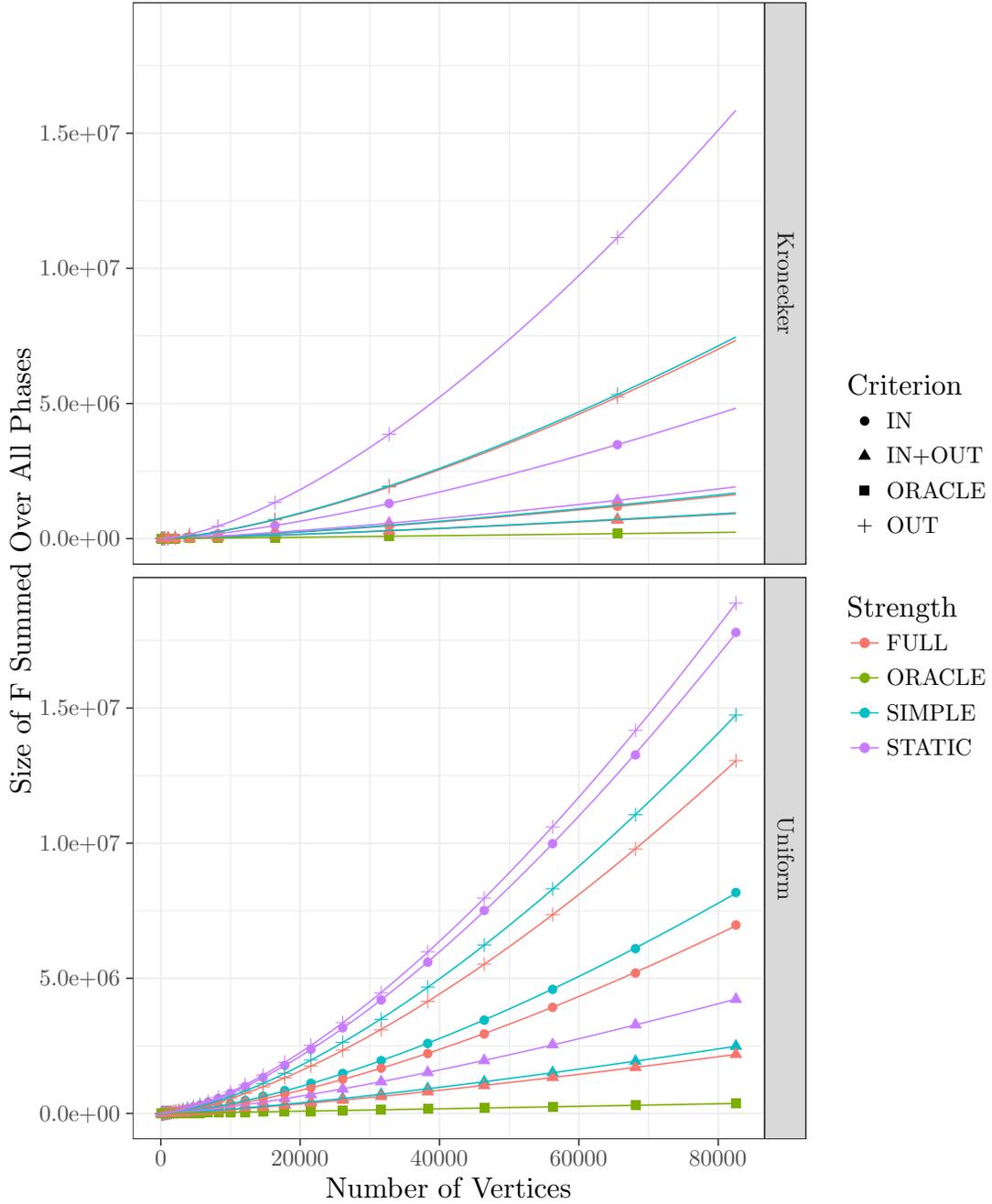}
    \caption{Sum of the sizes ~$|F|$ over all phases for each the
      criteria on uniform graphs with an expected out-degree of~10 for
      each vertex and Kronecker graphs with the initiator matrix
      $2.5\left(0.57 \ 0.19; 0.19 \ 0.05\right)$.}
    \label{simulation_f}
\end{figure}

Figure~\ref{simulation_f} shows the mean of the sum of the sizes
of~$F$ over all phases. This number can be used to estimate the amount
of work needed to settle all reachable vertices. As one can see
stronger criteria lead to a smaller $\sum|F|$, \ie, the reduction of
number of phases is so powerful that the sum becomes smaller. Again,
we used curve-fitting to to obtain the results seen in
Table~\ref{f_curve_fitting}. We fitted the function $a + b \cdot n^c$
with the parameters $a$, $b$, and $c$.  Additionally, we tried to fit
$a + b \cdot n\log_2(n)$, which unfortunately did not fit the
empirical data well. The sum ranges for the criteria from~$n\sqrt{n}$
to~$n\sqrt[3]{n}$ while for the oracle the sum is almost linear with
respect to the number of vertices~$n$.

\begin{table}[]
  \centering
  \caption{Sum of the sizes of~$F$ over all phases for each the
    criteria on uniform graphs with an expected out-degree of~10 for
    each vertex and Kronecker graphs with the initiator matrix
    $2.5\left(0.57 \ 0.19; 0.19 \ 0.05\right)$.}
  \label{f_curve_fitting}
  \begin{tabular}{r|c|c}
    \textbf{Criterion} & \textbf{Uniform Graphs} & \textbf{Kronecker Graphs} \\
    \hline
    $\OUTSTATIC(v)$ & $0.79 \cdot n^{1.50}$ & $0.48 \cdot n^{1.53}$ \\
    $\INSTATIC(v)$ & $0.73 \cdot n^{1.50}$ & $0.5 \cdot n^{1.42}$ \\
    $\OUTSTATIC(v)\vee\INSTATIC(v)$ & $1.17 \cdot n^{1.33}$ & $0.7 \cdot n^{1.31}$ \\
    \hline
    $\OUTSIMPLE(v)$ & $0.64 \cdot n^{1.50}$ & $0.52 \cdot n^{1.45}$ \\
    $\INSIMPLE(v)$ & $0.39 \cdot n^{1.49}$ & $0.46 \cdot n^{1.33}$ \\
    $\OUTSIMPLE(v)\vee\INSIMPLE(v)$ & $0.95 \cdot n^{1.31}$ & $0.63 \cdot n^{1.26}$ \\
    \hline
    $\OUT(v)$ & $0.58 \cdot n^{1.49}$ & $0.49 \cdot n^{1.46}$ \\
    $\IN(v)$ & $0.35 \cdot n^{1.49}$ & $0.41 \cdot n^{1.34}$ \\
    $\OUT(v)\vee\IN(v)$ & $0.96 \cdot n^{1.29}$ & $0.55 \cdot n^{1.27}$ \\
    \hline
    $\ORACLE(v)$ & $2.49 \cdot n^{1.05}$ & $1.16 \cdot n^{1.08}$ \\
  \end{tabular}
\end{table}

Additionally, we simulated four graphs from the SNAP dataset~\cite{snapnets}.

\begin{description}
  \item[Web Graph \say{Berk Stan}] A directed web-graph consisting of~%
  685 thousand vertices and~7.6 million edges.
  \item[Web Graph \say{Notre Dame}] A directed web-graph consisting of~%
  325 thousand vertices and~1.5 million edges.
  \item[Road Network Texas] An undirected graph representing the
  road network of Texas, with~1.3 million vertices and~1.9 million edges.
  \item[Road Network Pennsylvania] An undirected graph representing
  the road network of Pennsylvania, with~1 million vertices and~1.5
  million edges.
\end{description}

Since our implementation is only able to work with directed graphs,
we preprocessed the two road networks in such a way that for each
edge~$(a, b)$ an additional edge~$(b, a)$ was inserted into the
input file. This means that the number of edges for the two road
networks has been doubled. Additionally, since the input graphs
are unweighted, we added a uniformly random edge weight between~0
and~1 for each edge. Using unweighted graphs would trivialize the
SSSP.

\begin{table}[]
  \centering
  \caption{Number of phases required for four simulated SNAP graphs:~%
  Web Graph Berk Stan, Web Graph Notre Dame, Road Network Texas,
  and Road Network Pennsylvania.}
  \label{snap_table}
  \begin{tabular}{r|c|c|c|c}
    \textbf{Criterion} & \textbf{Berk Stan} & \textbf{Notre Dame} & \textbf{TX} & \textbf{PA} \\
    $n$ & $685\,000$ & $325\,000$ & $1\,300\,000$ & $1\,000\,000$ \\
    \hline
    $\OUTSTATIC(v)$ & 6165 & 2350 & 32948 & 25938 \\
    $\INSTATIC(v)$ & 5029 & 2224 & 32904 & 26027 \\
    $\OUTSTATIC(v)\vee\INSTATIC(v)$ & 2289 & 875 & 28938 & 22811 \\
    \hline
    $\OUTSIMPLE(v)$ & 3114 & 1643 & 20046 & 15784 \\
    $\INSIMPLE(v)$ & 3762 & 1358 & 31710 & 25062 \\
    $\OUTSIMPLE(v)\vee\INSIMPLE(v)$ & 1622 & 692 & 18930 & 14903 \\
    \hline
    $\OUT(v)$ & 2454 & 1183 & 16261 & 12798 \\
    $\IN(v)$ &  2341 & 1040 & 27962 & 22118 \\
    $\OUT(v)\vee\IN(v)$ & 1365 & 601 & 16092 & 12740 \\
    \hline
    $\ORACLE(v)$ & 582 & 53 & 898 & 716
  \end{tabular}
\end{table}

\begin{figure}
    \centering
    \input{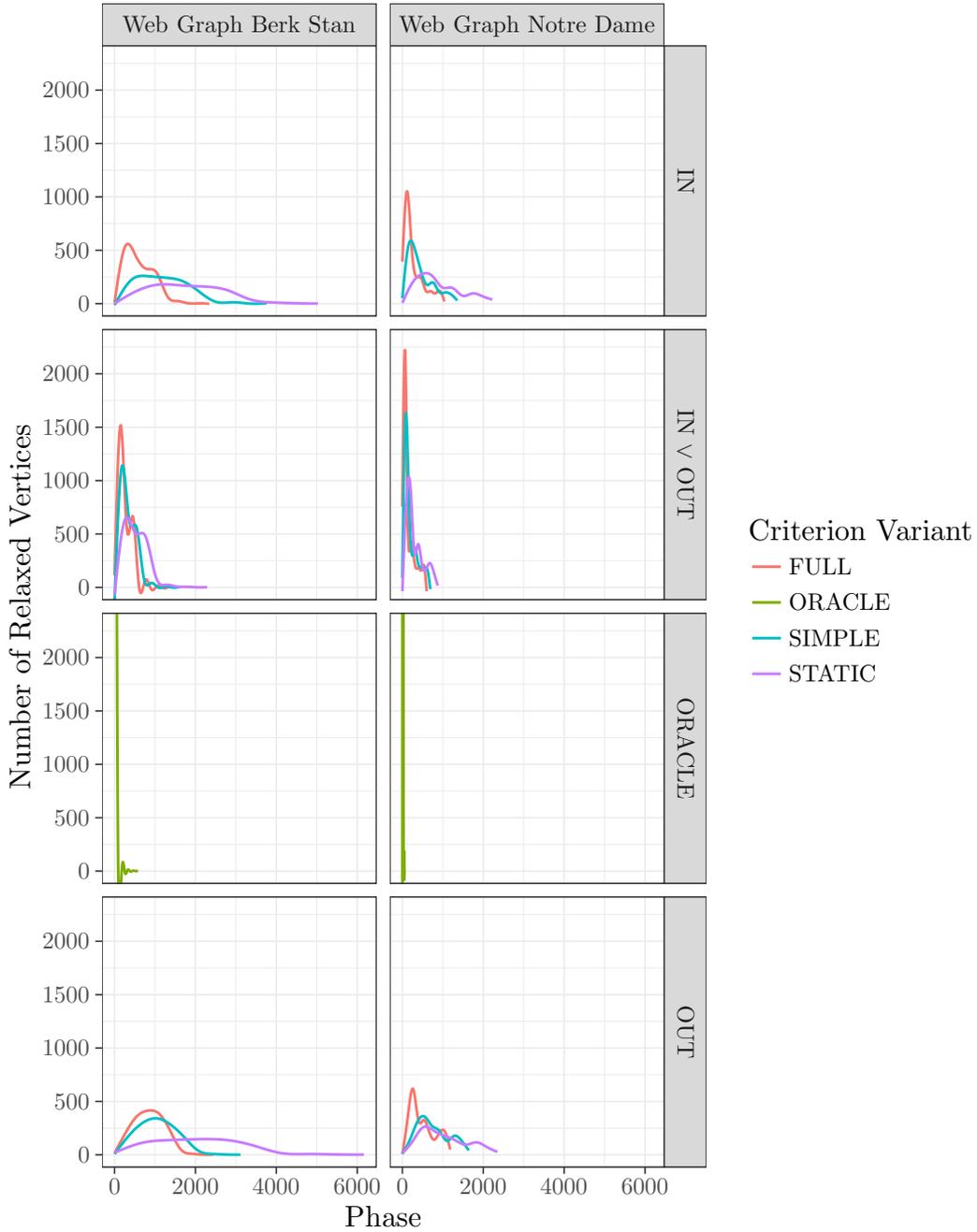}
    \caption{Number of vertices settled per phase for the
    two web graphs. The number of phases
    required can be seen by looking at the end of each line.
    The lines are smoothed in order to be able to
    display the data without heavy overplotting. The squiggly line
    at the end of $\ORACLE(v)$ is an artifact of this. $\ORACLE(v)$
    reaches up to~13000 for Berk Stan and up to~30000 for Notre Dame and had to
    be cut off in order to keep the graph legible.}
    \label{simulation_snap_web}
\end{figure}

Figure~\ref{simulation_snap_web} and Table~\ref{snap_table} show the results for the
two web graphs. One can see that neither $\IN(v)$ nor $\OUT(v)$
alone are able to realize the full potential of our approach
in reducing the number of phases. Only the combination $\IN(v)\vee\OUT(v)$
manages to do so. Different than for the road networks, for these
two graphs the difference between the static, simple and full
variations of our criteria is not as pronounced, especially
in the case of $\IN(v)\vee\OUT(v)$. $\ORACLE(v)$ still performed
an order of magnitude better than the strongest of our criteria.

\begin{figure}
    \centering
    \input{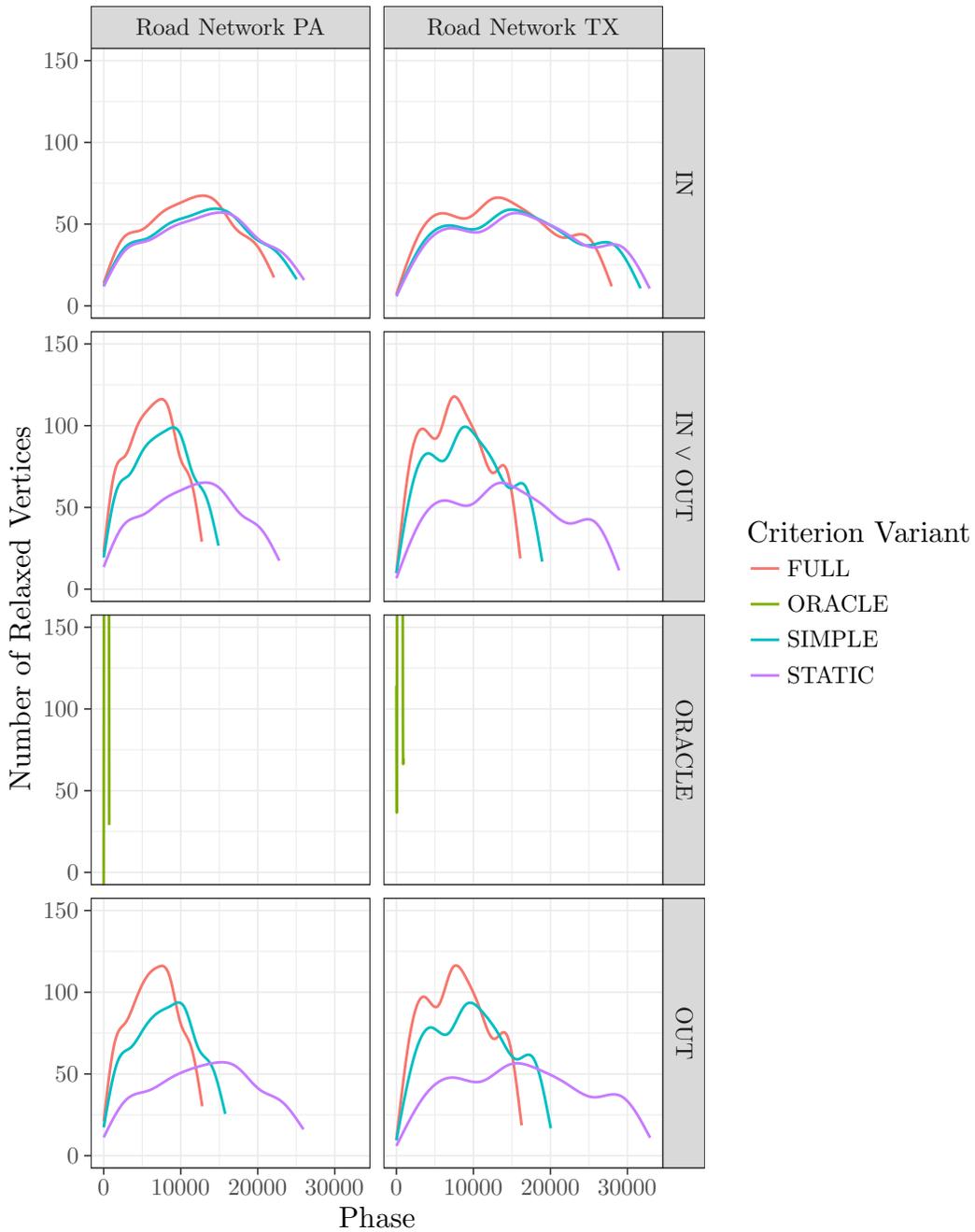}
    \caption{Number of vertices settled per phase for the
    two road networks. The number of phases
    required can be seen by looking at the end of each line.
    The lines are smoothed in order to be able to
    display the data without heavy overplotting. $\ORACLE(v)$
    reaches up to~3000 for PA and up to~4000 for TX and had to
    be cut off in order to keep the graph legible.}
    \label{simulation_snap_road}
\end{figure}

Figure~\ref{simulation_snap_road} and Table~\ref{snap_table} show the results for the
two road networks. One can see that for these two road networks
$\IN(v)$ is a quite weak criterion compared to $\OUT(v)$.
The combination of these two is not much stronger than $\OUT(v)$
alone. $\OUT(v)$ and $\OUTSIMPLE(v)$ manages to increase the
potential parallelism compared to $\OUTSTATIC(v)$, nevertheless
the settled vertices per phase are still low with about~100.
The theoretical optimum is much better than all our criteria.
It manages to settle about~3000 to~4000 thousand nodes per phase,
which implies that in these graphs there is still a lot of
untapped potential for much stronger criteria.

One can also see that the settling patterns are vastly different
between the two types of graphs:~The two road networks have a
steadily growing number of settled nodes, which after reaching its
peak steadily declines, \ie, there is potential for parallelism
in almost all phases. The two web graphs have a sharp increase
in the first few phases followed by a long tail
of phases where only very few vertices are settled. In this tail
there is hardly any potential for parallelism. This can
especially be seen in the Berk Stan graph.

\section{Implementations}
\label{sec:implementation}

We now describe our parallel, shared-memory implementations of the
SSSP algorithm running in phases and utilizing the $\INSTATIC(v)$ and
$\OUTSTATIC(v)$ criteria, and our implementation of the
$\Delta$-stepping algorithm.%
\footnote{%
The source code can be downloaded at \url{https://github.com/kaini/sssp-shm}.%
}
The implementations use a standard, adjacency-array
representation of the input graphs. A graph is stored as an array of
vertices and an array of edges.  The array of edges is grouped by the
source vertex of each edge, such that all outgoing edges of a vertex
are stored consecutively in memory. Each vertex is identified by its
index in the vertex array and consists of a pointer to the group of
its outgoing edges and the number of outgoing edges (outdegree).  For
the implementations here, the incoming edges of each vertex are not
strictly needed and storing them would roughly double the space
requirements.  Incoming edges are therefore not stored.

The static criteria have been implemented as proposed by Crauser et
al.. The tentative distances $d[v]$ are kept in a priority queue as in
Dijkstra's algorithm. For the $\INSTATIC(v)$ criterion an additional
priority queue is maintained that stores $d[v] - \min_{w\in V,
  (w,v)\in E}c(w,v)$ for each vertex, while for the $\OUTSTATIC(v)$
criterion an additional priority queue is used that stores $d[v] +
\min_{w\in V, (w,v)\in E}c(v,w)$ for each vertex. The initial minima
are computed for each vertex at the beginning of the
implementation. This takes $O(m)$ time and is included in the time
measurements of the next section to provide a fair comparison.  These
priority queues are then used to quickly identify the vertices for
which Equation~\eqref{crit:insimple}~(as long as $d[v] - \min_{w\in V,
  (w,v)\in E}c(w,v) \leq d[u]$ is true, the equation holds for $v$),
respectively Equation~\eqref{crit:outsimple}~(as long as $d[u] \leq
d[v] + \min_{w\in V, (w,v)\in E}c(v,w)$ is true, the equation holds
for $u$), hold. For the combination of the two criteria it suffices to
check for both conditions disjunctively.

Once a vertex has been identified by either criteria, it is removed
from all priority queues and the check is repeated.  When no criteria
identifies a vertex, all vertices that will be settled in the phase
have been collected. A sequential implementation would now just
iterate over the set of identified vertices and settle each of them as
done in Dijkstra's algorithm. Such an implementation would not have
any advantage over to Dijkstra's algorithm; it would probably be
slower because of the multiple priority queues that have to be
maintained instead of the single queue in Dijkstra's algorithm.

The parallel implementation is written in the \CC{} programming
language and uses native \CC{} threads. Additionally, we needed to implement
three primitives not provided by the \CC{} standard library:~a reduction
operation, an atomic-min operation, and a barrier.

The reduction, given a starting value~$s$, each processor's contribution~%
$c$ and an operation~$\oplus$, is implemented by utilizing a shared
atomic variable~$v$. Each processor reads the shared variable, calculates~%
$c \oplus v$ and tries to store the result back into~$v$ utilizing
a compare-and-exchange operation. If this fails, the whole process
is retried.
While this is a na\"ive implementation, the time required by the
reduction operations is completely irrelevant, and therefore does not
warrant implementing a more complex algorithm.

The atomic-min operation is implemented similarly,
with the small optimization that the compare-and-exchange does not
have to be retried if the own contribution is already greater or equal
to the read value.

The barrier implementation is heavily inspired by Boost's
barrier~\cite{boost}, but was reimplemented without using locks (mutex'es) and
condition-variables. A barrier consists of an atomic integer~$w$~(waiting)
and an atomic boolean~$g$~(generation). Once a processor enters
the barrier it performs an atomic fetch-and-increment operation
on~$w$ and fetches~$g$.
If after this operation~$w \neq p$, the processor busy-waits
until the value of~$g$ is flipped.
If after this operation~$w = p$, the processor flips~$g$ and
executes an atomic subtraction $w \coloneqq w - p$.

As a datastructure for all priority queues we utilized Boost's
Fibonacci heap implementation~\cite{boost} paired with a custom
allocator that avoids the negative performance impact of repeatedly
calling the system allocator on such an allocation-heavy
datastructure. We also tried to use Pairing heaps, but they turned out
to be slower than using Fibonacci heaps.

The implementation can be split into three separate phases:

\begin{enumerate}
  \item Preprocessing: Calculate the minimum outgoing/incoming edge
    for each vertex.
  \item Per-phase identification: Identify the vertices to be settled
    in this phase according to the criteria.
  \item Per-phase settling: Settle the identified vertices, relax
    tentative distances and update the priority queues.
\end{enumerate}

In our parallel implementation, each of these phases are parallelized.
We assume we have a set of $p$ processors (cores), each running a
thread.  Our implementation statically partitions the set of vertices
over the processors such that each processor (thread) is responsible
for a statically assigned subset of vertices. Each processor performs
all operations related to these vertices, and no processor performs
update operations on vertices assigned to another processor. In our
current implementation the assignment is not randomized, that is
processor $i$ is assigned vertices $v$ for which $v/p=i$.

\paragraph{Preprocessing:}
Calculating the cheapest outgoing edge for each vertex is trivial:
Each processor just iterates over the outgoing edges of vertices it is
responsible for.  Calculating the cheapest incoming edge for each
vertex is a bit more involved since processors do not know the
incoming edges of their vertices. We use a global array of~$n$ atomic
doubles, initialized to~$\infty$. Each processing unit iterates over
the outgoing edges of the vertices it is responsible for and uses an
atomic-min operation to update the cell corresponding to the target
vertex of the edge with the cost of the edge. When all processors have
finished doing so, each can read the cost of the cheapest incoming
edge from this array for the vertices it is responsible for.

\paragraph{Identification:}
To calculate the set of vertices identified by $\INSTATIC(v)$ we use a
priority queue ordered by $d[v] - \min_{w\in V, (w,v)\in
  E}c(w,v)$. Since each processor is only aware of the vertices it is
responsible for, each manages an independent priority queue containing
only vertices it is responsible for.  To find the minimum required to
decide $\INSTATIC(v)$, each processor first finds the minimum from its
own priority queue. Second, a reducing operation is executed across
all processors using these minima to obtain a global minimum tentative
distance. Once each processor knows the global minimum, they can now
independently identify the vertices that can be settled among these
they are responsible for. The vertices satisfying $\OUTSTATIC(v)$ are
found for each processor in the same way.

\paragraph{Settling:}
Once the vertices to be settled in the current phase are identified,
the relaxation is executed for each outgoing edge of these vertices,
with each processor relaxing vertices it is responsible for. For each
relaxation it is decided whether it is local, meaning that the target
vertex is belonging to the same processor, or whether it is remote,
meaning that the target vertex belongs to some other processor. Local
relaxations are executed immediately. For global relaxations, the
target vertex and new tentative distance is buffered in an array owned
by the destination processor. Once all processors are done iterating
over all outgoing edges of the vertices to be settled, they iterate
over all buffered relaxations they received in their array and execute
them. Settling is complete when all processors have finished this step.
In order to keep the update work small, remote relaxations are buffered
only if the can potentially improve the tentative distance of the target vertex.
To this end an approximate set of tentative distances is stored in an array
and updated with an atomic store by processors whose updates improve the
previously stored value. It does not matter that this array is not accurate,
therefore expensive atomic operations can be avoided here.

The buffers for incoming relaxations are implemented by using
an array of chunks and a counter variable.
Each chunk contains~$1024^2$~(about one million) items. The array of
chunks is large enough that the theoretically maximum needed number of
chunks can be allocated. Initially, the array of chunks contains
only null-pointers and the counter is~0.
If a processor wants to place something into a buffer
it first executes an atomic fetch-and-increment operation on the
counter variable to obtain the index
the item may be placed in. The first chunk is responsible for indices~0
to~$1024^2 - 1$, the second for~$1024^2$ to~$2 \cdot 1024^2 - 1$,
and so on. Therefore, once a processor obtained the index it knows the
chunk and the index in the chunk. If the target chunk is still null,
the processor allocates memory for the chunk and places the pointer
into the array of chunks using an atomic compare-and-exchange operation.
If another processor allocated the memory first, the memory allocated
by the other processor is used. Once the chunk is allocated, it suffices
for all processors to just place their item at the index obtained by
the fetch-and-increment operation, \ie, almost always an insertion
into such a buffer consists of a single
atomic fetch-and-increment operation and a simple~(non-atomic) write into
an array.

There are~$p$ such buffers, one for each processor where all \emph{other}
processors place the \emph{incoming} relaxations. Therefore, iterating
over all incoming relaxations is very simple. Each processor just has
to iterate over its own buffer.

Our implementation of $\Delta$-stepping~\cite{MeyerSanders03} follows
the same principles: Each processor maintains its own set of buckets,
\ie, each a bucket for vertices whose tentative distance is between~0
and~$\Delta$, between~$\Delta$ and~$2\Delta$, and so on. To find the
bucket that has to be emptied in any given iteration each processor
proposes its first non-empty local bucket. By reduction operation over
the proposed buckets, the globally first non-empty bucket is
identified. Each processor then empties this bucket and performs all
relaxations for light edges~(edges whose cost is less
than~$\Delta$). If the bucket is non-empty after this step it is
repeated. Finally, once all processors have completed all light edge
relaxations, in potentially several repetitions of the previous step, the
remaining heavy edges~(edges whose cost is larger than~$\Delta$) are
relaxed. Similarly to our implementation of Crauser's algorithm, local
relaxations are executed immediately, while remote relaxations are
buffered in an array for the destination processor. This means that all
processors have to wait for each other after each phase, and after each
iteration concerning the light edges. Once all buckets are empty, the
algorithm is finished. This can be easily detected by the reduction
operation that determines the globally first non-empty bucket.

\section{Experimental results}
\label{sec:results}

The first set of benchmarks was performed on a shared memory system~%
\say{mars} with eight Intel Xeon~E7-8850 processors.
Each processor has~10 cores and is capable of running~20 threads
in parallel. Each core has a base frequency of~2~GHz. The system
has about 1~TiB of main memory, but since our benchmarks
do not need a lot of memory, this does not matter.

All measurements were repeated at least~10 times, and the median
of these run-times was taken as basis for the following results.
The run-time of a single repetition
was obtained by using the maximum thread run-time.
In the case of uniformly random
and Kronecker graphs, each repetition used a different seed,
\ie, a different graph instance. Nevertheless, care was taken
that all criteria got the same set of different seeds to
ensure that the comparison stays absolutely fair.

Each benchmark was run in two configurations: First, the
criteria were implemented using Fibonacci-heaps as described in the
previous section. Second, each criterion was implemented using a
single plain array that is scanned linearly instead of utilizing
priority queue data-structures.
On one hand, this means that instead of find-min
operations, a simple linear scan was performed to find the minimum.
On the other hand, this also means that the performance overhead
of maintaining the priority queues goes away.

Each criteria is compared with an efficient sequential
implementation of Dijkstra's algorithm, \ie, the graphs
show the \emph{absolute speedup}. Our implementation of
Dijkstra's algorithm is made efficient by the fact that
we utilize Fibonacci-heaps with a hand-written custom
allocator that avoids the performance overhead of a heavily
allocating data-structure such as Fibonacci-heaps. Our implementation
of Dijkstra's algorithm is included in the source-code repository
linked in the previous section.

\begin{figure}
    \centering
    \input{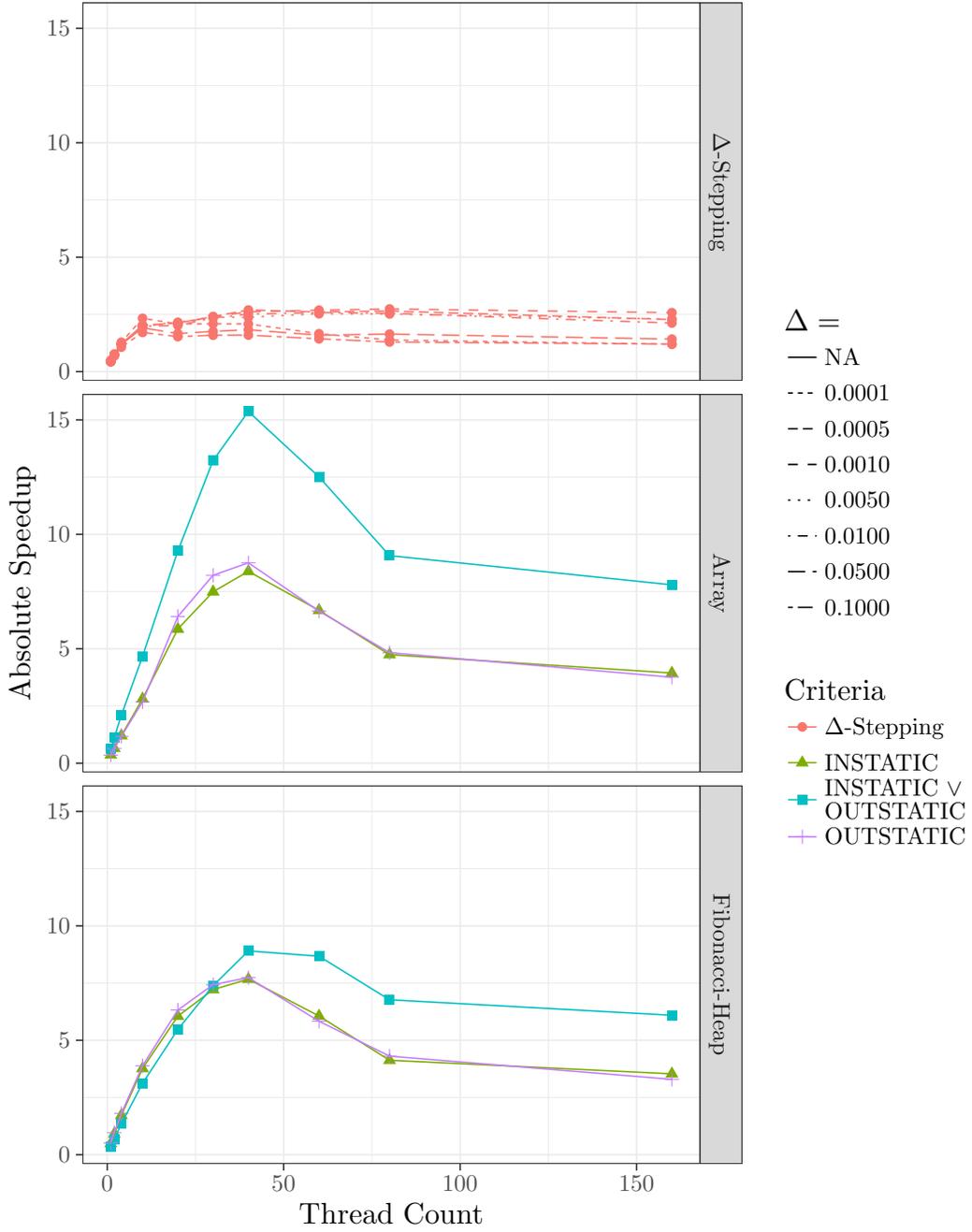}
    \caption{Absolute speedup of the $\INSTATIC(v)$ and
      $\OUTSTATIC(v)$ criteria, and $\Delta$-Stepping compared with an
      efficient implementation of Dijkstra's algorithm.
      The input graph is~$G(1000000, 0.0001)$ with uniformly random
      edge weights in~$[0; 1]$.
      The system used is mars.}
    \label{benchmark_uniform_mars}
\end{figure}

\begin{figure}
    \centering
    \input{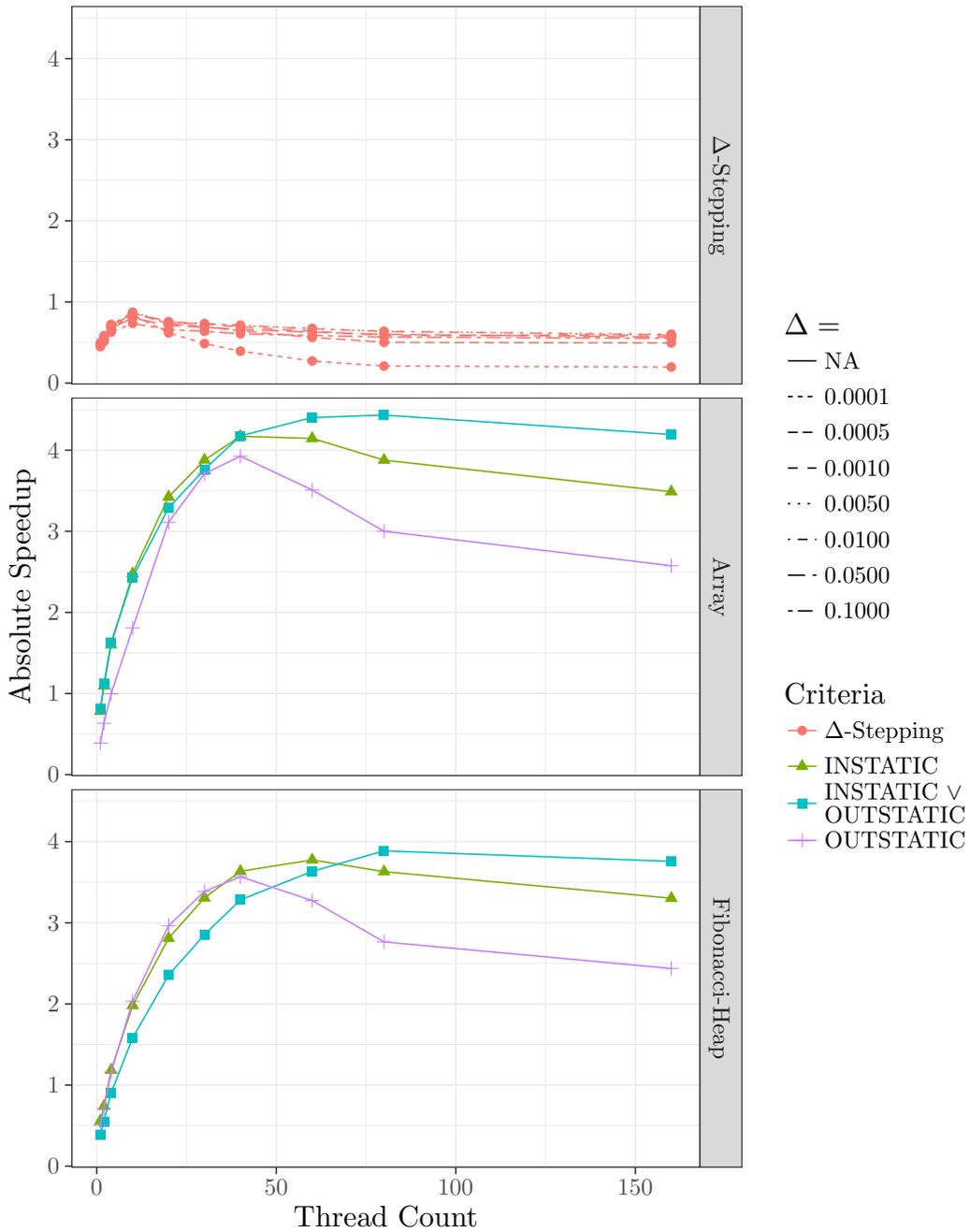}
    \caption{Absolute speedup of the $\INSTATIC(v)$ and
      $\OUTSTATIC(v)$ criteria, and $\Delta$-Stepping compared with an
      efficient implementation of Dijkstra's algorithm.
      The input graph is a Kronecker graph based on
      $\left(2.5\left(0.57 \ 0.19; 0.19 \ 0.05\right)\right)^{20}$
      with uniformly random
      edge weights in~$[0; 1]$.
      The system used is mars.}
    \label{benchmark_kronecker_mars}
\end{figure}

The first benchmark was performed on uniformly random graphs~%
$G(1000000, 0.0001)$ with uniformly random edge weights between~0
and~1. Therefore, each graph instance has exactly one million vertices
and about~100 outgoing edges per vertex, \ie, about~100 million edges
in total.  Figure~\ref{benchmark_uniform_mars} shows that the static
criteria as defined by Crauser et al.\ are indeed highly competitive
compared to $\Delta$-stepping.  We achieve an absolute speedup of up
to~15 when utilizing the combination of $\INSTATIC(v)$ and
$\OUTSTATIC(v)$ for identifying correct vertices. Unfortunately, the
algorithm seems to stop scaling for more than~40 threads on this
system.

For Kronecker graphs~(Figure~\ref{benchmark_kronecker_mars})
with uniformly random edge weights between~0 and~1 the
performance is much worse, as we only reach an absolute speedup
just shy of~4.5 with~80 threads. This seems to imply that the
structure of Kronecker graphs is indeed vastly different than
the structure of uniformly random graphs, in such a way that
it is much harder to achieve good speedups with our algorithm.

\begin{figure}
    \centering
    \input{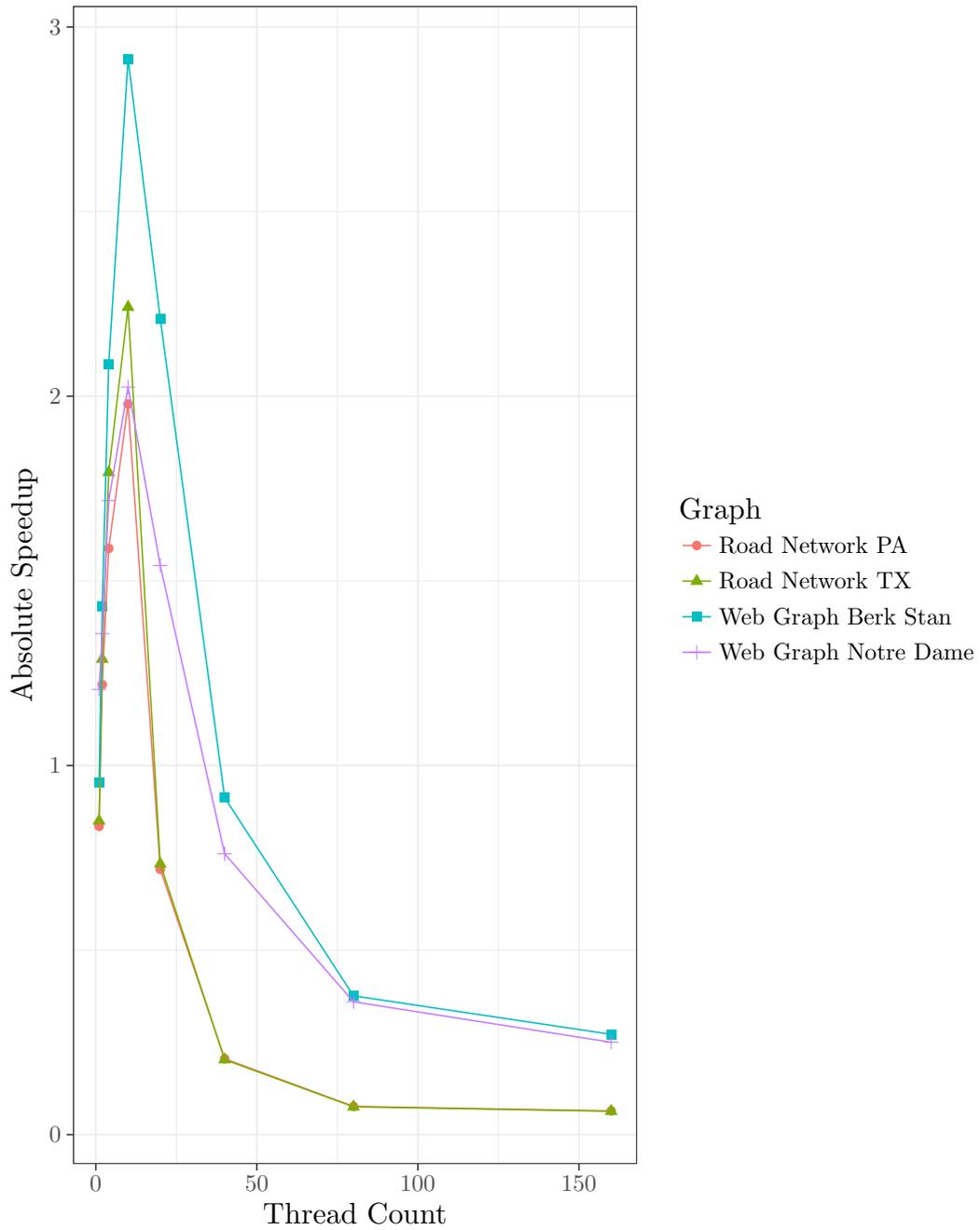}
    \caption{Absolute speedup of the $\INSTATIC(v) \vee \OUTSTATIC(v)$
      criterion using arrays, compared with an
      efficient implementation of Dijkstra's algorithm.
      The input graphs are from the SNAP dataset~\cite{snapnets} with
      uniformly random edge weights between~0 and~1.
      The system used is mars.}
    \label{benchmark_snap_mars}
\end{figure}

According to the previous results using
$\INSTATIC(v) \vee \OUTSTATIC(v)$ without priority queues but with
plain arrays instead is the fastest implementation.
Using this implementation, we performed benchmarks utilizing the
four graphs from the SNAP dataset introduced in the simulation,
\ie, the two web-graphs \say{Berk Stan} and \say{Notre Dame} and
the two~(preprocessed) road networks for Texas and Pennsylvania,
all with uniformly random edge weights between~0 and~1.
The results can be seen in Figure~\ref{benchmark_snap_mars}.
Unfortunately, these instances do not scale very well. We believe
that this is
due to the small size of the input graphs~(the edge counts range from~%
1.5 to~7.5 million).

\begin{figure}
    \centering
    \input{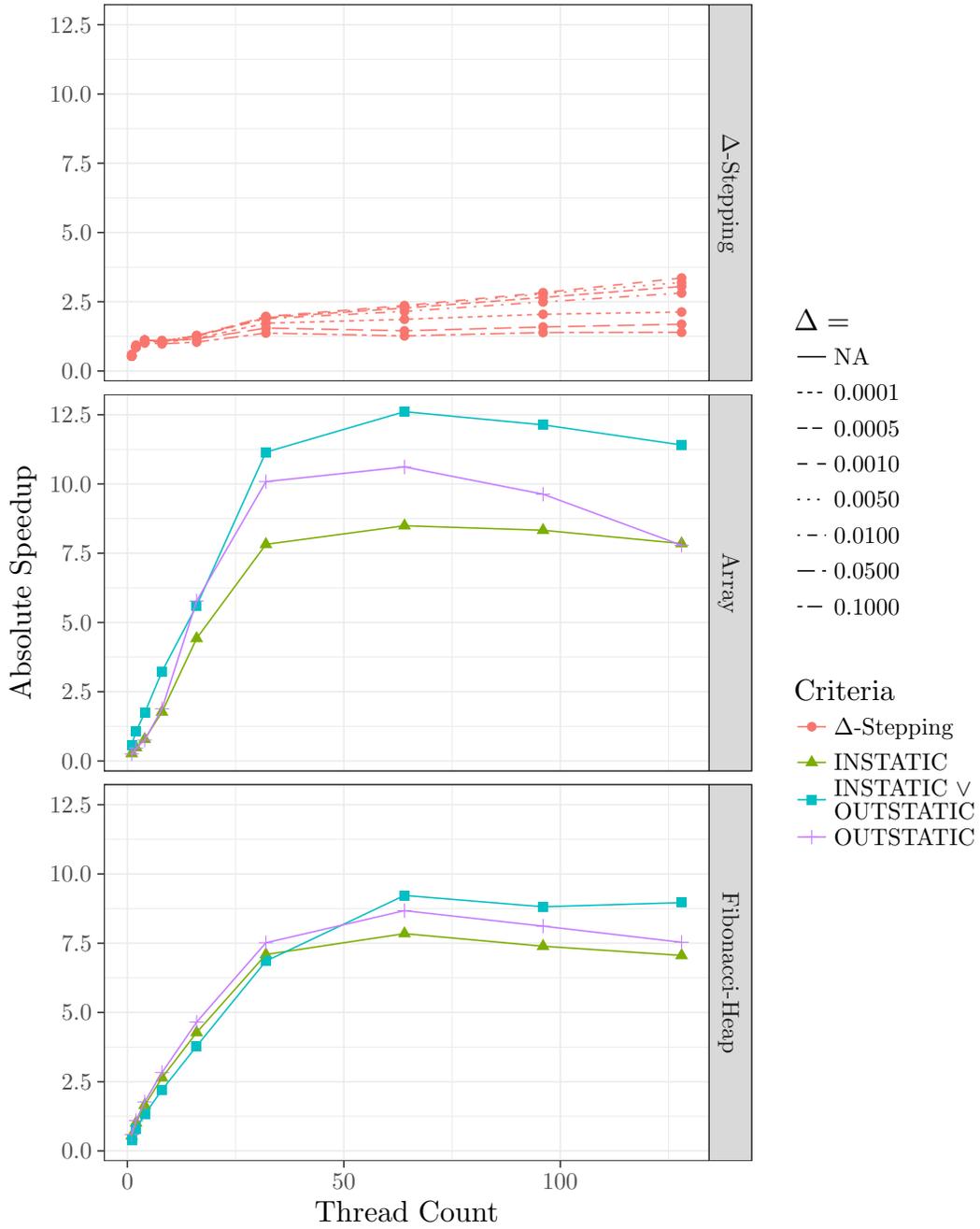}
    \caption{Absolute speedup of the $\INSTATIC(v)$ and
      $\OUTSTATIC(v)$ criteria, and $\Delta$-Stepping compared with an
      efficient implementation of Dijkstra's algorithm.
      The input graph is~$G(1000000, 0.0001)$ with uniformly random
      edge weights in~$[0; 1]$.
      The system used is nebula.}
    \label{benchmark_nebula}
\end{figure}

Additionally, we ran the first benchmark, i.e., random
graphs~$G(1000000, 0.0001)$ with uniformly random edge
weights in~$[0; 1]$ on a different system \say{nebula.}
This system consists of two AMD~EPYC~7551 CPUs, with
each a base clock speed of~2~GHz and 32 cores/64 threads.
The system has~256~GiB of main memory. The results can be
seen in Figure~\ref{benchmark_nebula}. The results are similar
to those of the system mars, with the notable exception that
on nebula the algorithms do not see a sharp decline in
absolute speedup for utilizing a high amount of threads.

We were not able to find an efficient implementation of the
stronger criteria discussed in Section~\ref{sec:criteria}.  A simple
implementation of $\INSIMPLE(v)\vee\OUTSIMPLE(v)$ led to speedups
worse than these of $\Delta$-stepping and is not usable in
practice. Nevertheless, it is included in the source-code package
linked in the previous section.

\section{Concluding remarks}

This paper discussed parallelization of Dijkstra's algorithm based on
criteria for detecting more than a single, correct candidate vertex for
relaxation at a time. We strengthened criteria previously introduced
by Crauser et al.~\cite{CrauserMehlhornMeyerSanders98}, and discussed
various ideas that can be used for practical implementation of
these. Simulation results show that for random and Kronecker graphs,
as often used in such studies, the (stronger) criteria can indeed
reduce the number of phases and therefore the parallel depth
significantly to a small root of the number of vertices in the input
graph. Stronger criteria indeed lead to stronger reduction in the
number of phases. We implemented a generic, parallel Dijkstra
algorithm running in phases, exploiting the two static criteria also
proposed by Crauser et al., and showed that for random graphs this
implementation can indeed be more than competitive with the
$\Delta$-stepping approach which is often claimed to be the fastest
and most efficient practical parallel SSSP implementation.

The implementations and discussions provided here leave much room for
further (practical) improvements, \eg, on the need for complex data
structures (priority queues), tradeoffs in the implementations
between criteria accuracy and overhead, etc. Based on the encouraging
speed-up results also in comparison to $\Delta$-stepping, we believe
that this is worthwhile.

\subsection*{Acknowledgments}

Thanks for discussions following the first revision of this note with
Ulrich Meyer, University of Frankfurt, and Vijay K.\ Garg, University of
Texas, Austin.

\FloatBarrier
\bibliographystyle{plain}
\bibliography{moredijkstra}

\end{document}